\documentclass[ALICE,manyauthors]{cernphprep}

\usepackage[comma,square,numbers,sort&compress]{natbib}
\usepackage{hyperref}
\usepackage{lineno}
\usepackage{subfigure}
\usepackage[section]{placeins}
\usepackage {multicol}
\usepackage {multirow}
\usepackage {units}
\usepackage {array}
\usepackage {xcolor}
\usepackage{comment}
\usepackage[T1]{fontenc}
\usepackage{orcidlink}

\newcommand{\fzero}{$\mathrm{f}_{0} (980)$}
\newcommand{\kstar}{$\mathrm{K}^{*}$(892)$^{0}$}
\newcommand{\rhoz}{$\rho$(770)$^{0}$}

\newcommand{\snn}          {\ensuremath{\sqrt{s_{\rm NN}}}}

\begin{document}

\begin{titlepage}

\PHyear{2023}
\PHnumber{263}      
\PHdate{16 November}  
%

\title{Observation of abnormal suppression of \textbf{f$_{0}$}(980) production \\in p--Pb collisions at $\sqrt{\textit{s}_{\mathbf{NN}}}$ = 5.02~TeV }

\ShortTitle{}   

\Collaboration{ALICE Collaboration\thanks{See Appendix~\ref{app:collab} for the list of collaboration members}}
\ShortAuthor{ALICE Collaboration} 

\begin{abstract}
The dependence of $\mathrm{f}_{0}$(980) production on the final-state charged-particle multiplicity in p--Pb collisions at $\sqrt{s_{\mathrm{NN}}} = 5.02$~TeV is reported. The production of $\mathrm{f}_{0}$(980) is measured with the ALICE detector via the $\mathrm{f}_0 (980) \rightarrow \pi^{+}\pi^{-}$ decay channel in a midrapidity region of $-0.5<y<0$. Particle yield ratios of $\mathrm{f}_{0}$(980) to $\pi$ and $\mathrm{K}^{*}$(892)$^{0}$ are found to be decreasing with increasing charged-particle multiplicity. The magnitude of the suppression of the $\mathrm{f}_{0}$(980)/$\pi$ and $\mathrm{f}_{0}$(980)/$\mathrm{K}^{*}$(892)$^{0}$ yield ratios is found to be dependent on the transverse momentum $p_{\mathrm{T}}$, suggesting different mechanisms responsible for the measured effects. Furthermore, the nuclear modification factor $Q_{\mathrm{pPb}}$ of $\mathrm{f}_{0}$(980) is measured in various multiplicity ranges. The $Q_{\mathrm{pPb}}$ shows a strong suppression of the $\mathrm{f}_{0}$(980) production in the $p_{\mathrm{T}}$ region up to about 4~GeV/$c$. The results on the particle yield ratios and $Q_{\mathrm{pPb}}$ for $\mathrm{f}_{0}$(980) may help to understand the late hadronic phase in p--Pb collisions and the nature of the internal structure of $\mathrm{f}_{0}$(980) particle.

\color{black}

\end{abstract}
 
\end{titlepage}

\setcounter{page}{2}


\section{Introduction}

Light scalar mesons, whose spin and parity are zero and even, respectively, are of particular interest as their nature can be explained with an exotic structure~\cite{ParticleDataGroup:2022pth}. Among them, a long-standing puzzle is related to the quark composition of the \fzero~particle~\cite{ExHIC:2010gcb, Jaffe:1976ig, Maiani:2004uc}. The \fzero~is suggested to be either a conventional meson ($\rm{}q\bar{q}$)~\cite{Chen:2003za}, a compact tetraquark~\cite{Achasov:2020aun}, or a $\rm{}K\overline{K}$ molecule~\cite{Ahmed:2020kmp}. By comparing different observables in heavy-ion collisions with those in pp interactions, the structure of the \fzero~can be probed.

The theory of the strong interaction, quantum chromodynamics (QCD), predicts the formation of a state of strongly interacting matter, the so-called quark--gluon plasma (QGP), under the conditions of high temperature and high energy density reached in relativistic heavy-ion collisions. Many observations at the Large Hadron Collider (LHC) and the Relativistic Heavy Ion Collider (RHIC), such as collective flow~\cite{Bhalerao:2020ulk, ALICE:2019zfl, Adams:2005dq, Adcox:2004mh} and jet quenching~\cite{ALICE:2019qyj, ATLAS:2010isq, PHENIX:2010nlr}, which is also manifest in the suppression of the yield of high-momentum hadrons~\cite{ALICE:2019hno, PHENIX:2006ujp} due to in-medium partonic energy loss, contribute to the understanding of the QGP properties~\cite{Heinz:2000bk, ALICE:2022wpn}. Specifically, the nuclear modification factors for different particle species, defined as the ratio of the transverse momentum ($p_{\mathrm{T}}$) distributions measured in heavy-ion collisions to the corresponding yields in pp interactions scaled by the number of nucleon--nucleon collisions, show a strong modification of the $p_{\mathrm{T}}$ spectra in large collision systems due to the presence of the hot and dense QGP medium. However, the nuclear modification factors are measured to be close to unity in minimum bias (MB) proton--nucleus (pA) collisions for $p_{\mathrm{T}}>$~8~GeV/$c$~\cite{ALICE:2016dei}, indicating no substantial modification in pA collisions in the high $p_{\mathrm{T}}$ range. 

Another effect observed in pp and pA collisions at the LHC is a multiplicity-dependent enhancement of the production of strange hadrons relative to hadrons composed of up and down quarks, which is usually referred to as ``strangeness enhancement''~\cite{ALICE:2016fzo}. The measurement of particle yield ratios of \fzero~to $\pi$ and \kstar~can be helpful to examine whether the \fzero~yield is influenced by the strangeness enhancement, thus providing sensitivity to the strange quark content inside the \fzero~\cite{LHCb:2014ooi, LHCb:2014vbo}. Moreover, other features observed in heavy-ion collisions, such as the strong enhancement of the nuclear modification factors of baryons at intermediate $p_{\mathrm{T}}$ (2~$<p_{\mathrm{T}}<$~5 GeV/$c$)~\cite{Fries:2003vb, ALICE:2022wpn} compared to those of mesons and the baryon and meson grouping of the elliptic flow~\cite{Wang:2022det}, show an apparent dependence on the number of constituent quarks (NCQ)~\cite{Wang:2022det}, reflecting the formation of hadrons from the QGP via quark coalescence~\cite{Fries:2003vb}. Hence, measurements of \fzero~production in systems where a QGP may be created can help to constrain the number of quarks forming the \fzero.

Short-lived resonances, such as \rhoz~\cite{ALICE:2018qdv}, \kstar~\cite{ALICE:2019etb, ALICE:2016sak}, $\Sigma(1385)^{\pm}$~\cite{ALICE:2022zuc}, and $\Lambda$(1520)~\cite{ALICE:2018ewo} as well as \fzero, are good probes to study the properties of the system that results from the hadronization of a QGP~\cite{Bierlich:2021poz, Knospe:2015nva}. In the late stage of the evolution of the system formed in heavy-ion collisions, there are two relevant temperatures and corresponding timescales: the chemical freeze-out, when the inelastic interactions among the constituents are expected to cease, and the later kinetic freeze-out, when all (elastic) interactions stop~\cite{Song:1996ik}. Since the time interval between the chemical and the kinetic freeze-outs of the system ($\sim$~10~fm/$c$) is comparable with the lifetime of resonances~\cite{ALICE:2011dyt, ALICE:2019xyr}, their decay products can actively interact with the hadronic gas via rescattering whereas regeneration can occur from interactions between particle pairs in the hadron gas. These two processes are designated as hadronic interactions in this Letter. The hadronic interactions result in modifications of resonance yields. The modifications can be studied by comparing the yield of resonances with those of long-lived or ground-state particles~\cite{ALICE:2018pal}. Measurements of \rhoz$/(\pi^{+}+\pi^{-})$~\cite{ALICE:2018qdv} and \kstar$\rm{}/(K^{+}+K^{-})$~\cite{ALICE:2019etb, ALICE:2016sak} yield ratios are good examples to study the properties of the late hadronic phase after the chemical freeze-out. It is worth mentioning that the ratios of particles with the same strangeness can eliminate potential strangeness enhancement effects in the ratio. Recently, system-size-dependent modifications of particle yields are also observed in small collision systems~\cite{ALICE:2016sak, ALICE:2019etb}, suggesting that rescattering and regeneration may also occur in high-multiplicity pp and p--Pb collisions. These hadronic interactions depend on the hadronic cross section of the decay products inside the hadronic medium, the lifetimes of the resonance, and the duration of the hadronic phase. The suppression of resonance yields in the hadronic gas can be explained by rescattering dominating over regeneration. In addition, the final states of resonances decaying to $\pi\pi$, such as \fzero, \rhoz, and f$_{2}$(1270), are affected by the same cross section of pions and the medium, while the amount of hadronic interactions differs due to different lifetimes of these resonances. In this context, measuring the modification of the \fzero~yield may contribute to further understanding of the late hadronic phase.

In this Letter, multiplicity-dependent measurements of \fzero~production in p--Pb collisions at center-of-mass energy per nucleon--nucleon collision $\snn$ = 5.02~TeV are reported for the first time. The \fzero~is measured at midrapidity ($-0.5<y<0$) in 0~$<p_{\mathrm{T}}<$~8~GeV/$c$ for different multiplicity classes. In Sec.~\ref{sec:setup}, the experimental setup is described, while the reconstruction of \fzero\ and the relative corrections are explained in Sec.~\ref{sec:ana}. The study of systematic uncertainties for the measurement is reported in Sec.~\ref{sec:syst}. In Sec.~\ref{sec:results}, $p_{\mathrm{T}}$ spectra, particle yield ratios, the nuclear modification factors, and model comparisons are discussed. Finally, conclusions are outlined in Sec.~\ref{sec:summary}.

\label{sec:intro}


\section{Experimental setup}
\label{sec:setup}
The sample of MB p--Pb collisions at $\snn$~=~5.02~TeV used for the present analysis was recorded using the ALICE detector in 2016. Due to the different energies of the proton and lead beams, the center-of-mass reference system in p--Pb collisions is shifted in rapidity by $\Delta y_{\mathrm{cms}} =$~0.465 along the direction of the proton beam. In the following, the convention that $y$ stands for $y_{\mathrm{cms}}$ is used. The ALICE apparatus during the LHC Run 2 is described in detail in Ref.~\cite{Abelev:2014ffa}. The present analysis is carried out using the following detectors: the V0~\cite{ALICE:2013axi}, the Zero Degree Calorimeters (ZDC)~\cite{Cortese:2019nnv}, the Inner Tracking System (ITS)~\cite{ALICE:2010tia}, the Time Projection Chamber (TPC)~\cite{Alme:2010ke}, and the Time-Of-Flight (TOF)~\cite{Jacazio:2018slq}. 

The V0 detector consists of two arrays of scintillators located on both sides of the interaction point (IP), denoted as V0A and V0C, each made of 32 plastic scintillator strips, covering the full azimuthal angle within the pseudorapidity intervals $2.8 < \eta < 5.1$ and $-3.7 < \eta < -1.7$, respectively. Minimum bias p--Pb collisions are selected online by requiring a signal in both V0A and V0C detectors in coincidence with the LHC bunch crossing. The total charge deposited in the V0A on the Pb-going side is utilized to define the multiplicity classes. The collected MB sample corresponds to an integrated luminosity of 0.3~nb$^{-1}$~\cite{ALICE:2014gvw}. The ZDC detects nucleons emitted from the colliding nucleus by nuclear de-excitation processes or knocked out from wounded nucleons, the so-called “slow” nucleons. Two identical sets of ZDCs, each composed of a neutron (ZN) and a proton (ZP) calorimeter, are located at 112.5 m from the ALICE IP on both sides, covering very forward rapidity regions. The ZDC provides the least biased centrality selection in p--Pb collisions~\cite{ALICE:2014xsp}.
 
The primary vertex position is reconstructed using the measured track segments in the Silicon Pixel Detector (SPD)~\cite{Santoro2009:ALICESPD}, the innermost two layers of the ITS. The primary vertex position along the beam direction ($z_\mathrm{vtx}$) is required to be in $|z_\mathrm{vtx}|<10$~cm from the nominal interaction point ($z_\mathrm{vtx}=0$). The pileup is reduced by rejecting events with multiple reconstructed vertices with the additional requirement that the distance between the primary vertex and any additional reconstructed vertex is larger than 0.8~cm. In addition, an inconsistency between the number of track candidates in the ITS and clusters in the SPD is used to further reduce the pileup events~\cite{ALICE:2015olq}. After these selections, the probability of pileup events is expected to be about 0.1\% in the MB sample~\cite{ALICE:2017svf}. Charged particles are reconstructed down to $p_{\mathrm{T}}=$~0.15~GeV/$c$ in the pseudorapidity range $|\eta|<$~0.9 over the full azimuth with the TPC and the ITS detectors, which are located inside a large solenoidal magnet, providing a uniform magnetic field of 0.5~T directed along the beam axis. Particle identification (PID) can be performed with the TPC and TOF. The TPC measures specific ionization energy loss $\mathrm{d}E/\mathrm{d}x$ of charged tracks to separate particle species. The TOF is used for PID by measuring the flight time of charged particles from the primary vertex to the TOF.

\section{Data analysis}

The \fzero~resonances are reconstructed via the decay channel \fzero~$\rightarrow \pi^{+}\pi^{-}$, for which the branching ratio is reported to be B.R. = ($46\pm6$)\%~\cite{Stone:2013eaa}. The \fzero~candidates are built from pairs of charged tracks reconstructed in the ITS and TPC. The tracks and required to have $p_{\mathrm{T}}>$~0.15~GeV/$c$ and $|\eta|<$~0.8 for a uniform detector acceptance. The reconstructed tracks are required to satisfy the standard selection criteria, as reported in Ref.~\cite{ALICE:2022qnb}, to guarantee that only tracks with high quality are selected. To ensure good track momentum resolution, the reconstructed tracks are required to have crossed at least 70 readout pad rows (out of a maximum of 159) in the TPC and to have at least two associated hits in the ITS (out of a maximum of 6), with at least one in the SPD. Selection criteria, which are dependent on $p_{\mathrm{T}}$, are applied to the distance of closest approach of the track to the primary vertex in the transverse ($d_{xy}$) and longitudinal ($d_{z}$) directions, requiring $|d_{z}|<$~2~cm and $|d_{xy}|<$~(0.0105~$+$~0.0350~$\times p_{\mathrm{T}}^{-1.1})$~cm (with $p_{\mathrm{T}}$ in GeV/$c$), respectively, to suppress contamination from secondary charged particles originating from weakly decaying hadrons and interactions with the material.

The identification of charged pions is performed using the combined information of the TPC and TOF. The difference between the measured ionization energy loss and the expected value from a Bethe--Bloch parameterization obtained by assuming the particle is a pion is required to be within two standard deviations for the pion identification in the TPC. The difference between the measured flight time of the particle and the expected flight time for a pion is required to be within three standard deviations for the particle to be identified as a pion in the TOF. Tracks not having a signal associated in the TOF are identified using only the $\mathrm{d}E/\mathrm{d}x$ information from the TPC.

\label{sec:ana}
\begin{figure}[hbt!]
	\centering
	\subfigure{ \includegraphics[width=0.47 \textwidth]{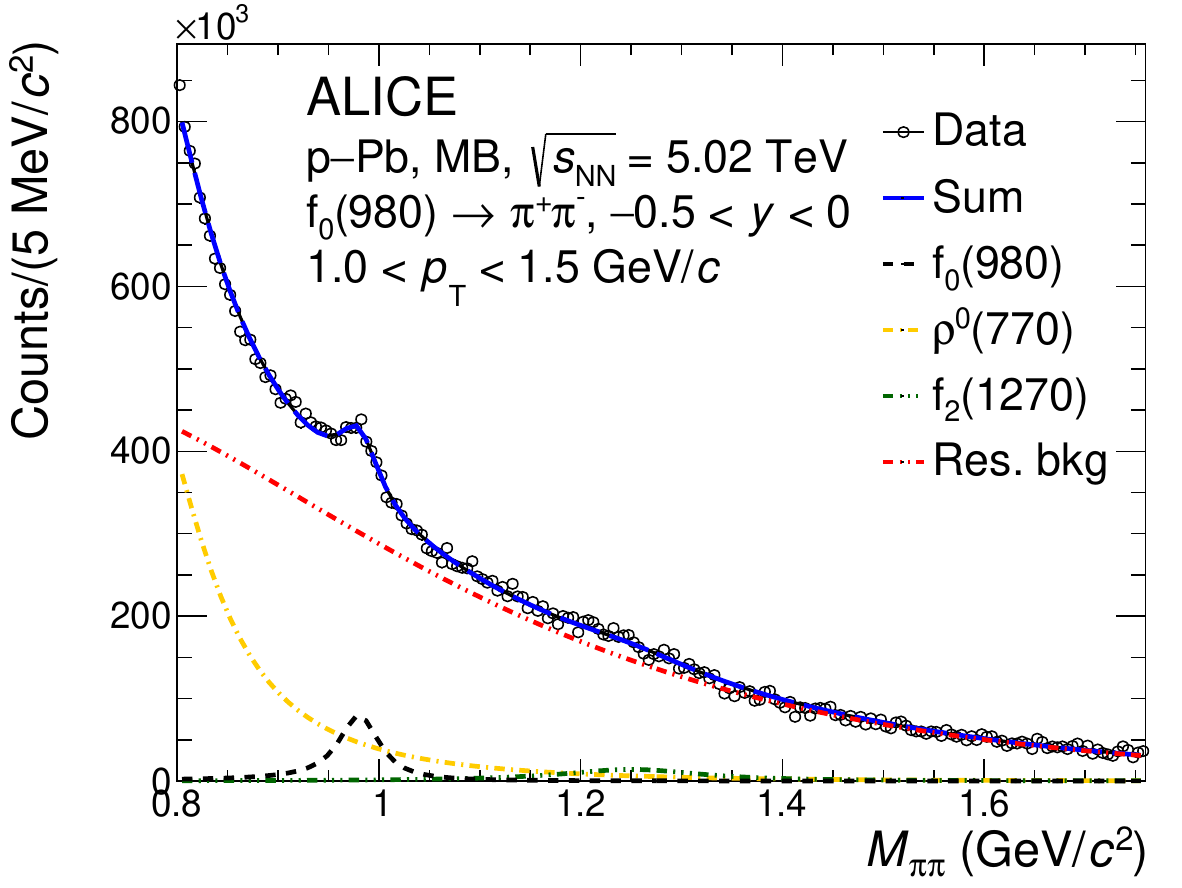} }
	\subfigure{ \includegraphics[width=0.47 \textwidth]{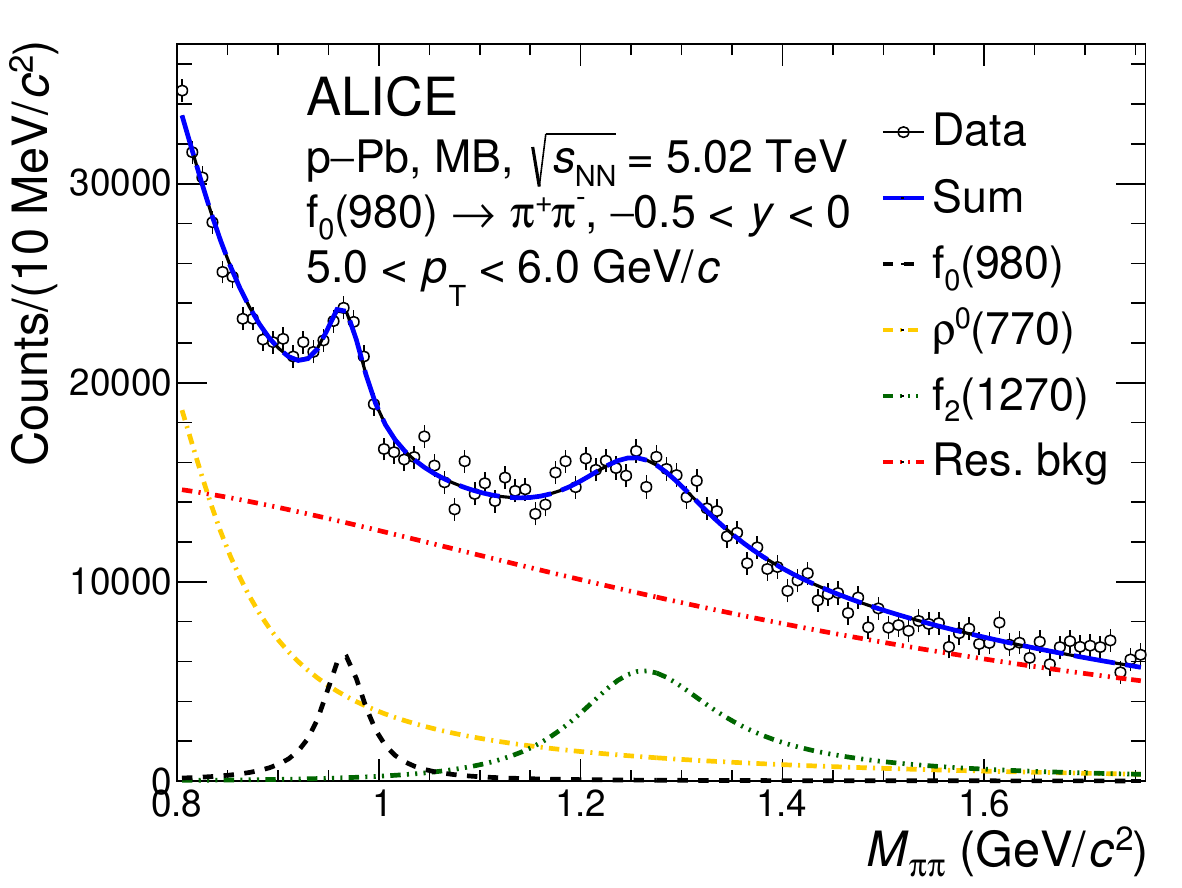} }
	\caption{ Invariant mass distribution of $\pi^{+}\pi^{-}$ pairs in $-0.5<y<0$ after the like-sign background subtraction in p--Pb collisions at \snn~=~5.02~TeV. The left (right) plot is obtained at low (high) $p_{\mathrm{T}}$ of $\pi^{+}\pi^{-}$ pairs in minimum bias events. The abbreviation of Res. bkg represents the residual background.}
	\label{fig:SigExt}
\end{figure}

The \fzero~signals are extracted using an invariant mass analysis by associating two opposite-charge pions in the same event within $-0.5<y<0$~\cite{ALICE:2013wgn}. The combinatorial background is subtracted using the like-sign method~\cite{LIKESIGN}. The like-sign background is constructed as the geometric average of $\pi^{+}\pi^{+}$ and $\pi^{-}\pi^{-}$ distributions, 2$\sqrt{N_{\pi^{+}\pi^{+}}N_{\pi^{-}\pi^{-}}}$. After subtracting the like-sign background from the $\pi^{+}\pi^{-}$ distribution, peaks of resonances decaying to $\pi^{+}\pi^{-}$ can be identified. Figure~\ref{fig:SigExt} shows the like-sign-subtracted $\pi^{+}\pi^{-}$ invariant mass ($M_{\pi\pi}$) distributions for 1.0~$<p_{\rm{T}}<$~1.5~GeV/$c$ (5.0~$<p_{\rm{T}}<$~6.0~GeV/$c$) in MB events in the left (right) panel. Because \rhoz~and $\rm{f}_{2}$(1270) dominantly decay to $\pi^{+}\pi^{-}$ and have large widths, \fzero~signals are overlapped with contributions from those two resonances. In addition, a residual background ($f_{\mathrm{bkg}}$) is present, which is mainly attributed to misidentified particles and mini-jets. The measured invariant mass distribution is fitted with a function accounting for the contributions of this residual background and of the three resonances. Each resonance contribution is described with a relativistic Breit-Wigner function (rBW)~\cite{ALICE:2018qdv, ALICE:2022qnb}. Note that the detector resolution of $\mathcal{O}$(1 MeV) gives a negligible contribution to the widths of broad resonances~\cite{ALICE:2016sak}. The rBW can be expressed as
\begin{eqnarray}
\mathrm{rBW}(M_{\pi\pi}) = \dfrac{AM_{\pi\pi}\Gamma(M_{\pi\pi})M_{0}}{(M_{\pi\pi}^{2}-M_{0}^{2})^{2} + M_{0}^{2}\Gamma^{2}(M_{\pi\pi})},
\label{eq:rBW}
\end{eqnarray}
where $\Gamma(M_{\pi\pi})$ is defined as
\begin{eqnarray}
\Gamma(M_{\pi\pi}) = \left[ \dfrac{ (M_{\pi\pi}^{2} - 4m_{\pi}^{2}) }{ (M_{0}^{2}-4m_{\pi}^{2}) } \right]^{(2J+1)/2} \times \dfrac{\Gamma_{0}M_{0}}{M_{\pi\pi}} .
\label{eq:rBWW}
\end{eqnarray}
Here, $A$ and $M_{0}$ are the amplitude of the rBW and the rest mass of the resonance, respectively. The rest width of the resonance, the spin, and the charged pion mass of 139.6~MeV/$c^{2}$ are represented as $\Gamma_{0}$, $J$, and $m_{\pi}$, respectively. The spins for \fzero, \rhoz, and $\mathrm{f}_{2}$(1270) are 0, 1, and 2, respectively. Each resonance rBW is corrected for the phase space factor~\cite{ALICE:2018qdv}, which can be expressed as
\begin{eqnarray}
\mathrm{PS}(M_{\pi\pi}) = \dfrac{M_{\pi\pi}}{\sqrt{M_{\pi\pi}^{2}+p_{\mathrm{T}}^{2}}}\times\exp{(-\sqrt{M_{\pi\pi}^{2}+p_{\mathrm{T}}^{2}}/T_{\mathrm{kin}})},
\label{eq:ps}
\end{eqnarray} 
where $p_{\mathrm{T}}$ denotes the transverse momentum of the $\pi\pi$ pair and is set to be the median of each $p_{\mathrm{T}}$ interval, and $T_{\mathrm{kin}}$ is the kinetic freeze-out temperature, set to be 160~MeV~\cite{ALICE:2018qdv} for all the defined multiplicity classes. The fit function for the background, $f_{\mathrm{bkg}}$, is modeled with a Maxwell-Boltzmann-like distribution, which can be expressed as~\cite{OPAL:1998enc}
\begin{eqnarray}
f_{\mathrm{bkg}}(M_{\pi\pi}) = B(M_{\pi\pi}-2m_{\pi})^{n}\exp{(c_{1}M_{\pi\pi} + c_{2}M_{\pi\pi}^{2})},
\label{eq:bkg}
\end{eqnarray} 
where, $B$, $n$, $c_{1}$, and $c_{2}$ are free parameters. 

The total fit function consists of the sum of three rBWs, one for each resonance and one function for the background, $f_{\mathrm{bkg}}$. This function has nine free parameters: three for \fzero~resonance (mass, width, and amplitude), two amplitudes for \rhoz~and f$_{2}$(1270) resonances, and four parameters for $f_{\mathrm{bkg}}$. In particular, the width of the \fzero~which is not yet constrained by measurements (10~$<\Gamma_{0}^{\mathrm{f}_{0}}<$~100~MeV/$c^{2}$~\cite{ParticleDataGroup:2022pth}) is left as a free parameter in the fit. The masses and widths of \rhoz~and $\mathrm{f}_{2}$(1270) are fixed to their world-average values from Ref.~\cite{ParticleDataGroup:2022pth}, namely $m_{\rho}=$~766.5~MeV/$c^{2}$, $\Gamma^{\rho}_{0}=$~~149.1~MeV/$c^{2}$, $m_{\mathrm{f}_{2}}=$~1,275.5~MeV/$c^{2}$, and $\Gamma^{\mathrm{f}_{2}}_{0}=$~186.7~MeV/$c^{2}$. Due to the many free parameters in the fit function, the procedure is split into three steps to prevent parameter values from converging to local minima. The purpose of the first step is to obtain an unbiased initial value for the \fzero~width. This step is performed using the MB sample over a coarse $p_{\mathrm{T}}$ binning to reduce the effect of statistical fluctuations. This coarse $p_{\mathrm{T}}$ binning is defined by merging 2 or 3 $p_{\mathrm{T}}$ bins of the finer $p_{\mathrm{T}}$ binning used for the analysis. All nine parameters are left free in the first step. The second step aims at constraining the $f_{\mathrm{bkg}}$. The \fzero~width is fixed to the value determined from the wider $p_{\mathrm{T}}$ interval used in the previous step. The last step is processed fixing the parameters of $f_{\mathrm{bkg}}$ to those extracted in the previous step, while the \fzero~width is allowed to vary in the range of 10~$<\Gamma_{0}^{\mathrm{f}_{0}}<$~100~MeV/$c^{2}$. In this procedure, the amplitudes of the three resonances and the mass of the \fzero~are left free, and the fit range is set to 0.8~$<M_{\pi\pi}<$~1.76~GeV/$c^{2}$. In the last step, the extracted width of \fzero~ranges between 40 and 70~MeV/$c^{2}$ in the different $p_{\mathrm{T}}$ and multiplicity intervals used in this analysis. The determination of the \fzero~width is sensitive to the modeling of the background and the other two resonances in the fit.

While for the \fzero~analysis performed in pp collisions~\cite{ALICE:2022qnb}, the width was constrained to be 55 MeV/$c^{2}$, the present analysis leaves the \fzero~width as a free parameter. In the previous analysis, no phase space correction was applied. On the other hand, the present analysis considers the phase space correction for a possibly larger probability of $\pi\pi$ interference~\cite{STAR:2003vqj} owing to higher multiplicity in p--Pb collisions. It is found that consistent invariant yields in pp collisions are obtained from the two different analysis methods.

The raw yields of \fzero~($N_{\mathrm{f}_{0}}$) in each $p_{\mathrm{T}}$ interval are obtained by integrating the \fzero~rBW function. They are corrected for the acceptance, the tracking efficiency, and the PID efficiency and then normalized for the number of selected p--Pb collisions, the width of the $p_{\mathrm{T}}$ and rapidity interval, and the B.R.~\cite{Stone:2013eaa}. The fully corrected yield can be expressed as
\begin{eqnarray}
\dfrac{1}{N_{\mathrm{NSD}}}\dfrac{\mathrm{d}^{2}N}{\mathrm{dyd}p_{\mathrm{T}}} = \dfrac{1}{N_{\mathrm{evt}}} \dfrac{ N_{\mathrm{f}_{0}} }{ \Delta \mathrm{y} \Delta p_{\mathrm{T}} } \dfrac{  \epsilon_{\mathrm{trig}} f_{\mathrm{vtx}} f_{\mathrm{SL}} }{\mathrm{Acc} \times \epsilon \times \mathrm{B.R.} }.
\end{eqnarray}
Here, the number of events satisfying the event selection criteria in the specific multiplicity class is represented as $N_{\mathrm{evt}}$. The corrected yield is then normalized to $N_{\mathrm{NSD}}$, which is the number of non-single diffractive (NSD) events, via the factors $\epsilon_{\mathrm{trig}} \times f_{\mathrm{vtx}} \times f_{\mathrm{SL}}$. The width of the rapidity interval (of 0.5 units) is represented as $\Delta y$. Coefficients for the acceptance ($\mathrm{Acc}$) and the efficiency ($\epsilon$) of the tracking and PID for pion pairs are estimated from a detailed simulation of the ALICE detector response. The p--Pb collisions are simulated using the DPMJET~\cite{Fedynitch:2015kcn} event generator with the injection of \fzero~signals. The generated particles (signal and background) are transported through the detector using GEANT3~\cite{Brun:1994aa}. The $\mathrm{Acc}\times\epsilon$ is estimated to be 26\% in the 0~$<p_{\mathrm{T}}<$~0.3~GeV/$c$ interval and gradually increasing up to 60\% as $p_{\mathrm{T}}$ increases, without any dependence on the multiplicity class. The $\mathrm{B.R.}$ is the branching ratio of the \fzero~$\rightarrow \pi^{+}\pi^{-}$ decay channel. The \fzero~yield is normalized for the trigger efficiency ($\epsilon_{\mathrm{trig}}$), vertex reconstruction efficiency ($f_{\mathrm{vtx}}$), and signal loss ($f_{\mathrm{SL}}$) due to the event selection. The $\epsilon_{\mathrm{trig}}$ depends on the multiplicity class increasing from 0.84 to 1 as the multiplicity increases. The $f_{\mathrm{vtx}}$ is estimated to be larger than 0.99 in all measured multiplicity classes. The $f_{\mathrm{SL}}$ corrects for the \fzero~signal loss due to the event selection. Because general Monte Carlo event generators do not generate primary \fzero~particles, the $f_{\mathrm{SL}}$ is estimated using a different particle, the $\phi$ meson, exploiting the universal $m_{\mathrm{T}}$ scaling~\cite{Altenkamper:2017qot}. This approach shows that $f_{\mathrm{SL}}$ does not depend on particle species~\cite{ALICE:2019xyr}, and it is found to be 1.03 for 0~$<p_{\mathrm{T}}<$~0.3~GeV/$c$ and approaching unity for $p_{\mathrm{T}}>$~2~GeV/$c$.

\section{Systematic uncertainties}
\label{sec:syst}
The systematic uncertainties of the \fzero~yields are estimated by varying the analysis selection criteria, the configuration of the fit used to extract the raw yield, and the treatment of the phase space correction. The estimated uncertainties are summarized in Table~\ref{tab:syst}. The total systematic uncertainty is calculated as the quadratic sum of the different contributing sources. The estimated uncertainties are different in the different multiplicity classes and the $p_{\mathrm{T}}$ intervals, but they do not show a clear trend as a function of multiplicity and $p_{\mathrm{T}}$. In Table~\ref{tab:syst}, the minimum and maximum uncertainty values are reported for each source. The relative uncertainty of the $\mathrm{B.R.}$ is 13\%~\cite{Stone:2013eaa} and is not included in the total uncertainty.

\begin{table}[h!]
\caption{Relative systematic uncertainties of the \fzero~$p_{\rm{T}}$-differential yields. Numbers given in ranges correspond to minimum and maximum uncertainties.}
\centering
\begin{tabular}{l|c}
\hline 
Sources &Systematic uncertainty (\%) \\ \hline
Primary vertex selection & negligible \\ 
Pileup rejection & negligible \\ 
Tracking & 4 \\
Particle identification & 4--13 \\ 
$\mathrm{f}_{2}$(1270) parameters	& 2--10 \\ 
\rhoz~parameters & 2--9 \\
Fit range & 0--7 \\
Initial $\mathrm{f}_{0}$ width & 2--13 \\
Phase space correction & 2--9 \\ \hline 
Total (in quadrature)	& 11--23 \\ 
\hline 
\end{tabular}
\label{tab:syst}
\end{table}

The systematic uncertainty from the primary vertex selection is estimated by repeating the analysis with a different selection of $|z_\mathrm{vtx}|<$~7~cm and found to be negligible. The systematic uncertainty from the pileup rejection is tested by varying the minimum number of track segments contributing to the reconstruction of pileup collision vertices from 5 to 3. The uncertainty is estimated to be negligible.

The systematic uncertainty from tracking is taken from~\cite{ALICE:2013wgn}, where uncertainties are evaluated by varying the requirements to select reconstructed primary tracks such as those on $d_{xy}$, $d_{z}$, and the number of crossed rows. The estimated uncertainty is 4\%. The systematic uncertainty from the PID is tested with different requirements on the number of standard deviations ($\pm\,0.5\sigma$ with respect to the default selections) for the TPC and TOF selection. The uncertainties are estimated to range from 4\% to 13\% in the different $p_{\mathrm{T}}$ intervals and multiplicity classes.

The uncertainties due to the \fzero~yield extraction via the fits to the invariant mass distributions are estimated by varying some of the configurations in the fit procedure. The contributions coming from masses and widths of $\mathrm{f}_{2}$(1270) and \rhoz~are evaluated by shifting the masses and the widths by three standard deviations with respect to their world-average values using the uncertainties reported in Ref.~\cite{ParticleDataGroup:2022pth}. The estimated systematic uncertainties are 2--10\% and 2--9\%, respectively. Furthermore, the invariant mass range used in the fit is changed inward or outward by 40~MeV$/c^{2}$, and the resulting systematic uncertainty is found to be less than 7\%. The contribution from the initial guess for the \fzero~width value, which is obtained in the first fit step described in Sec.~\ref{sec:ana}, is estimated by varying the width within the statistical uncertainties, which is about 5~MeV/$c{^2}$ on average, in both directions. The variations affect the background distribution determined in the second step, and the estimated systematic uncertainties range from 2\% to 13\%. The systematic uncertainty from the phase space correction is estimated by varying the kinetic freeze-out temperature in the range of 140~$<T_{\mathrm{kin}}<$~180~MeV. The estimated uncertainties range from 2\% to 9\%. 

The correlations of the systematic uncertainties in different multiplicity classes are quantified. The uncertainty is considered more closely correlated when the directions of the systematic deviations are the same for different multiplicity classes. This is tested by comparing the directions of the systematic deviations between a given multiplicity class and the MB class. For all the sources of systematic uncertainty it is found that approximately half of the total systematic uncertainties are uncorrelated.


\section {Results}
\label{sec:results}

\begin{figure}[!hbt]
	\centering
	\subfigure{ \includegraphics[width=0.6 \textwidth]{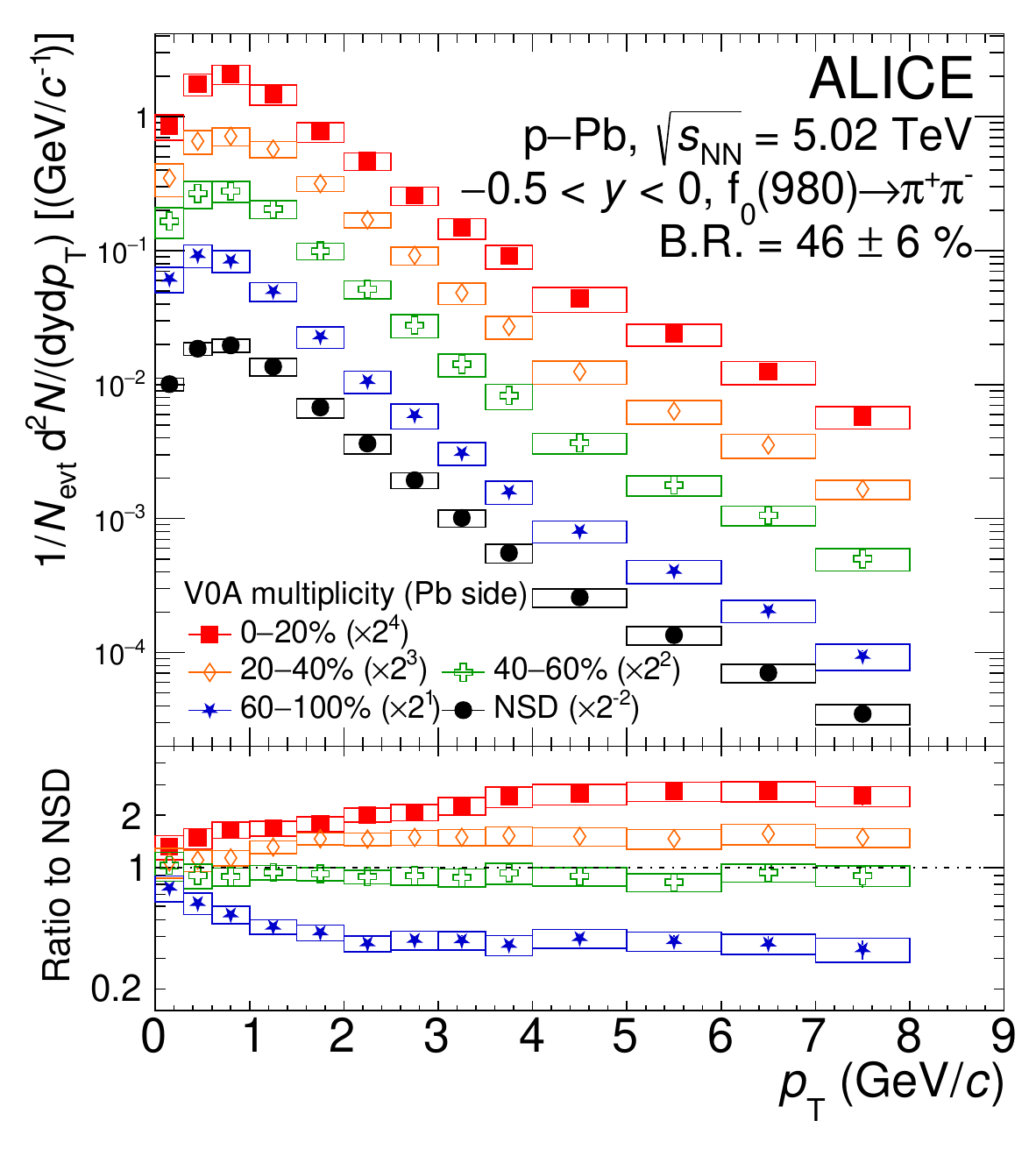} }
	\caption{ Transverse momentum spectra of \fzero~in p--Pb collisions at \snn~=~5.02~TeV for different multiplicity classes, which are scaled for visibility. Statistical and systematic uncertainties are shown as error bars and boxes, respectively. The normalization uncertainty of 13\% due to the uncertainty of the $\mathrm{B.R.}$ is not shown in the figure. The lower panel shows the ratios of the spectra in multiplicity classes to the NSD spectrum. }
	\label{fig:pt}
\end{figure}

Figure~\ref{fig:pt} shows the $p_{\mathrm{T}}$ spectra of \fzero~in p--Pb collisions at \snn~=~5.02~TeV measured in the range of 0~$<p_{\mathrm{T}}<$~8~GeV/$c$ for different multiplicity classes and NSD events. The multiplicity classes are defined based on the V0A amplitudes, which are proportional to the multiplicity of particles in the forward rapidity region of the Pb-going side. Each spectrum is scaled by a multiplicative factor denoted in the figure for visibility. The lower panel of Fig.~\ref{fig:pt} shows the ratios of the $p_{\mathrm{T}}$ spectra in different multiplicity classes to the NSD one. The systematic uncertainties of the ratios are estimated by propagating the multiplicity-uncorrelated uncertainties on the individual spectra. For $p_{\mathrm{T}}<$~4~GeV/$c$, a hardening of the $p_{\mathrm{T}}$ spectrum from low- to high-multiplicity events is clearly seen, while the spectral shapes in the different multiplicity classes are found to become consistent among each other for $p_{\mathrm{T}}>$~4~GeV/$c$. Such trends are similar to those observed for other hadronic species~\cite{Schnedermann:1993ws, ALICE:2019hno} and are understood as due to the radial flow.

\begin{table}[h!]
\caption{The values of d$N$/d$y$ and $\left\langle p_{\mathrm{T}} \right\rangle$ for \fzero~measured in p--Pb collisions at \snn~=~5.02~TeV for different multiplicity classes~\cite{ALICE:2012xs}. The first and second uncertainties represent the statistical and systematic uncertainties, respectively}
\centering
\begin{tabular}{ccc}
\hline 
Multiplicity class (V0A) & d$N$/dy & $\left\langle p_{\mathrm{T}} \right\rangle$ (GeV/$c$) \\ \hline
0--20\% & 0.206$\pm$0.005$\pm$0.014 & 1.287$\pm$0.034$\pm$0.010 \\
20--40\% & 0.153$\pm$0.004$\pm$0.010 & 1.250$\pm$0.029$\pm$0.082 \\
40--60\% & 0.113$\pm$0.002$\pm$0.008 & 1.142$\pm$0.025$\pm$0.088 \\
60--100\% & 0.064$\pm$0.001$\pm$0.005 & 0.999$\pm$0.014$\pm$0.080 \\
\hline
\end{tabular}
\label{tab:ymp}
\end{table}

The integrated yield (d$N$/d$y$) and mean $p_{\mathrm{T}}$ $\left( \left\langle p_{\mathrm{T}} \right\rangle \right)$ of \fzero~are calculated by integrating and averaging the $p_{\mathrm{T}}$ spectrum, respectively. Table~\ref{tab:ymp} shows the d$N$/d$y$ and $ \left\langle p_{\mathrm{T}} \right\rangle$ of \fzero~for different multiplicity classes in p--Pb collisions at \snn~=~5.02~TeV. The d$N$/d$y$ of \fzero~is found to increase linearly with charged-particle multiplicity when considering the $\left\langle \mathrm{d}N_{\mathrm{ch}}/\mathrm{d}\eta \right\rangle$ values from Ref.~\cite{ALICE:2012xs}, while the $\left\langle p_{\mathrm{T}} \right\rangle$ exhibits a weak dependence on the charged-particle multiplicity. 

\begin{figure}[!hbt]
	\centering
	\includegraphics[width=0.49 \textwidth]{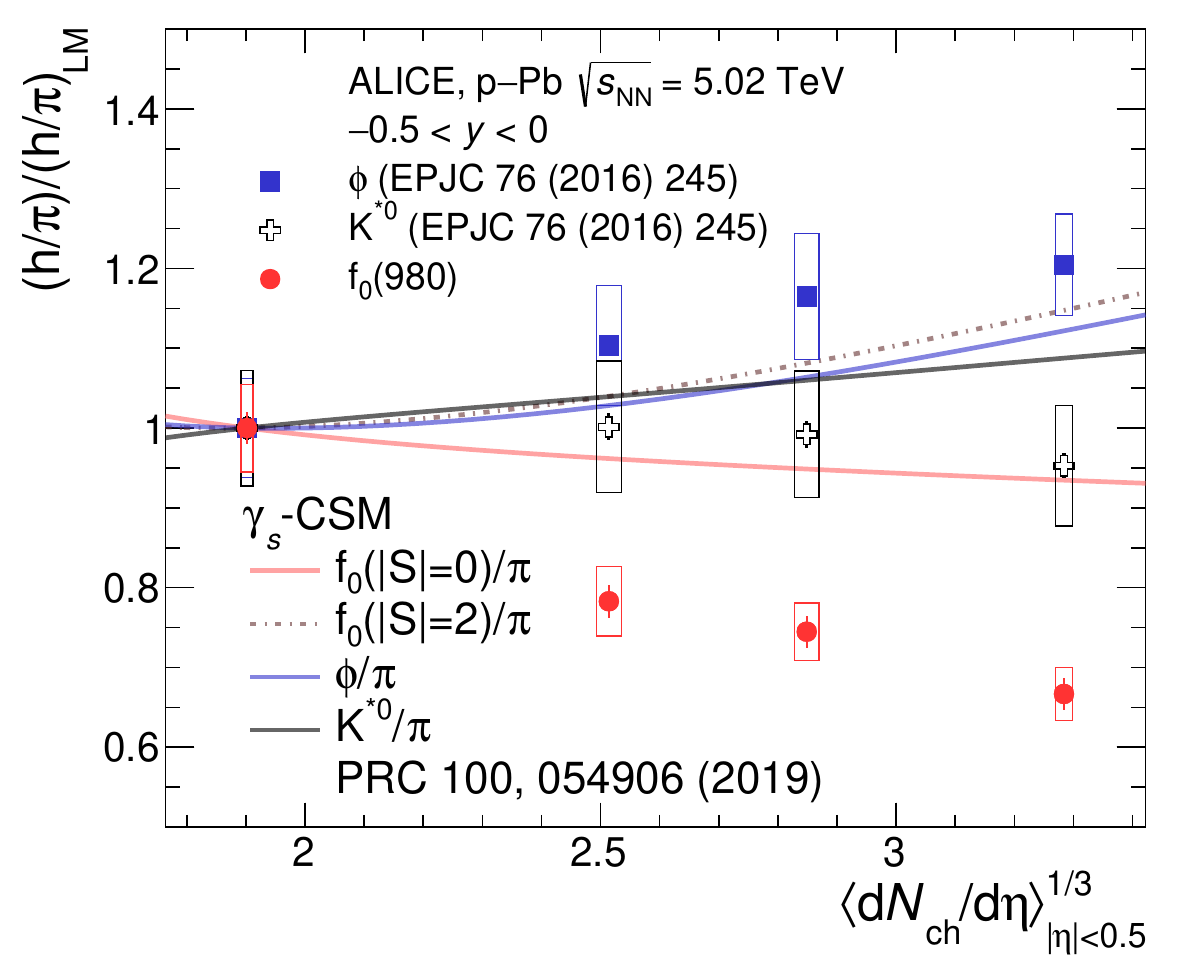} 
        \includegraphics[width=0.49 \textwidth]{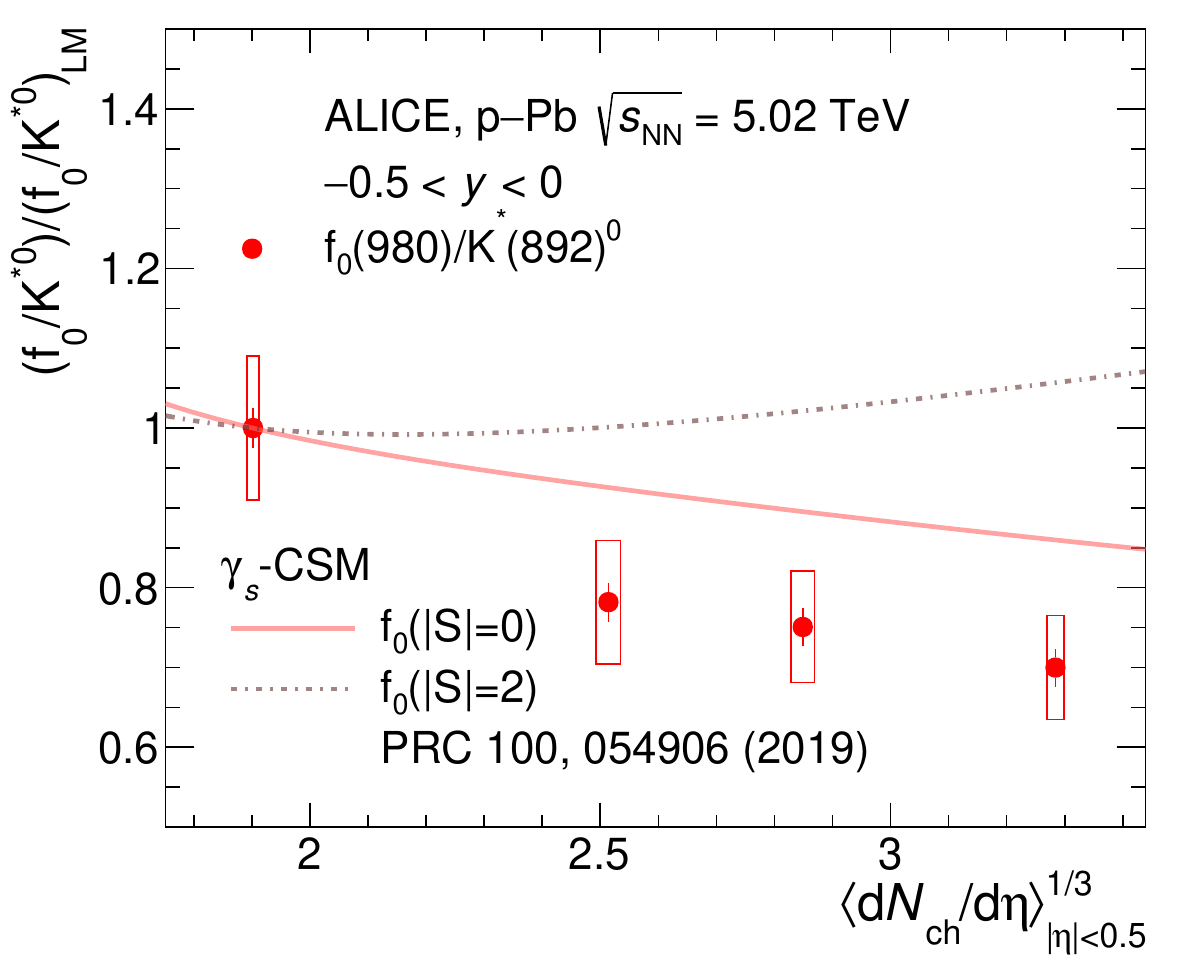} 
	\caption{ Double ratios of $\phi$~\cite{ALICE:2016sak}, \kstar~\cite{ALICE:2016sak}, and \fzero~to $\pi$~\cite{ALICE:2016dei} (left) and \fzero~to \kstar~(right) as a function of charged-particle multiplicity raised to the power of 1/3. The ratio in each multiplicity class is divided by the ratio in low-multiplicity (LM, 60--100\%) events to allow for a direct comparison among different hadron species and reduce systematic uncertainties. V0A is utilized to categorize events based on their multiplicity. Predictions from the canonical statistical model are represented with lines. }
	\label{fig:f0piAddCSM}
\end{figure}

The multiplicity dependence of \fzero~production is studied by comparing the ratio of the yields of different hadron species to that of pions in different multiplicity classes. A double ratio is calculated by dividing the hadron-to-pion yield ratios to their values measured in the lowest multiplicity interval (60--100\%), $(h/\pi)/(h/\pi)_{\mathrm{LM}}$, allowing a direct comparison of multiplicity dependence among different hadron species and the reduction of the systematic uncertainties. The left panel of Fig.~\ref{fig:f0piAddCSM} shows the double ratios of different particles to charged pion yields as a function of charged-particle multiplicity raised to the power of 1/3 in p--Pb collisions at \snn~=~5.02~TeV. The $\langle \mathrm{d}N/\mathrm{d}\eta \rangle_{|\eta|<0.5}^{1/3}$ is a proxy for the size of the system~\cite{Liu:2018xae}. The systematic uncertainty of the double ratio is calculated considering only the uncorrelated part of the uncertainty. The pion~\cite{ALICE:2016dei}, \kstar~\cite{ALICE:2016sak}, and $\phi$~\cite{ALICE:2016sak} mesons can be classified according to their lifetimes and to their (anti-)strange quark content. The strangeness enhancement and hadronic interactions can be studied by comparing the yield of particles with different characteristics. The double ratio of $\phi$ to $\pi$ increases with increasing multiplicity, which is consistent with the effect of the strangeness enhancement~\cite{ALICE:2016fzo}. Due to the long lifetime ($\approx$~46.2~fm/$c$) of $\phi$ meson, little effect is expected from interactions in the hadronic phase. On the other hand, the double ratio of \kstar~to $\pi$ is independent of multiplicity within the uncertainties even if \kstar~contains one strange quark. The flat trend could be explained by two competing effects, the strangeness enhancement and the interactions of the decay particles in the hadronic medium, due to the short lifetime ($\approx$~4.2~fm/$c$) of \kstar~\cite{ParticleDataGroup:2022pth}. One can expect that hadronic interactions reduce the \kstar~yield if the rescattering dominates over the regeneration. The double ratio of \fzero~to $\pi$ decreases as the multiplicity increases because of the short lifetime ($\approx$~3--5~fm/$c$ from the $\Gamma_{0}^{\mathrm{f}_{0}}$ range estimated from our fits to the \fzero~line shape) of \fzero, suggesting that rescattering effects may play a role. Predictions of the ratio of \fzero~to $\pi$ are shown in Fig.~\ref{fig:f0piAddCSM} (left) for different hidden strangeness ($|S|$) assumptions for \fzero~in the $\gamma_{s}$-Canonical Statistical Model (CSM)~\cite{Vovchenko:2019kes}, where $|S|$ is the number of strange and anti-strange quarks. The CSM considers system-size-dependent hadrochemistry at vanishing baryon density with local conservation of electric charge, baryon density, and strangeness while allowing for undersaturation of strangeness. Note that $\gamma_{s}$ is the parameter for the undersaturation of strangeness and is derived from a fit to the measured particle yields. The CSM hypothesis with two hidden strange quarks predicts an increase of the double ratio, contrary to what is observed experimentally. Moreover, the CSM with zero hidden strangeness predicts the \fzero/$\pi$ ratio to decrease much less than what is measured. When comparing the predicted trend to the measured one, it should however be considered that the CSM does not model interactions in the hadronic phase. The prediction of the CSM for the $\phi/\pi$ ratio qualitatively reproduces the increasing trend of the data with increasing multiplicity, where the hidden strangeness of $\phi$ is two. However, the CSM overestimates the ratio of \kstar~to $\pi$ at high multiplicity because the modification of \kstar~yields due to rescattering effects is not implemented in the CSM, while the strangeness enhancement for \kstar~is included.

The right panel of Fig.~\ref{fig:f0piAddCSM} shows the double ratio of the \fzero~to \kstar~yield as a function of $\left\langle \mathrm{d}N_{\mathrm{ch}}/\mathrm{d}\eta \right\rangle^{1/3}$ together with the predictions from the CSM with different hidden strangeness assumptions. The lifetimes of \fzero~and \kstar~are estimated to be of similar order of magnitude and both smaller than the duration of the hadronic phase in p--Pb collisions~\cite{ParticleDataGroup:2022pth}. This leads to the expectation that the \fzero/\kstar~ratio is weakly affected by hadronic interactions, which depend on the hadronic cross section of the different decay products of the two resonances. The measured double ratio shows a decreasing trend with increasing multiplicity, which is qualitatively described with the zero-hidden-strangeness assumption for \fzero~and can be explained by the strangeness enhancement of the \kstar~yield. The CSM prediction with the assumption of two hidden strange quarks is mildly increasing as the multiplicity increases, a trend that is opposite to the experimental result. The differences between the data point at the highest multiplicity and the two predictions with zero and two strange quarks amount to 2.3 and 5.2 standard deviations, calculated using the total uncertainty on the measurement, respectively. Therefore, the decreasing trend of the double ratio of \fzero~to \kstar~can suggest no effective strangeness enhancement for the \fzero. The discrepancies between the data and the model predictions can be attributed to the assumption of the same modification of \fzero~and \kstar~yields from hadronic interactions while the decay products from \fzero~and \kstar~differ.

\begin{figure}[!hbt]
	\centering
	\subfigure{ \includegraphics[width=0.6 \textwidth]{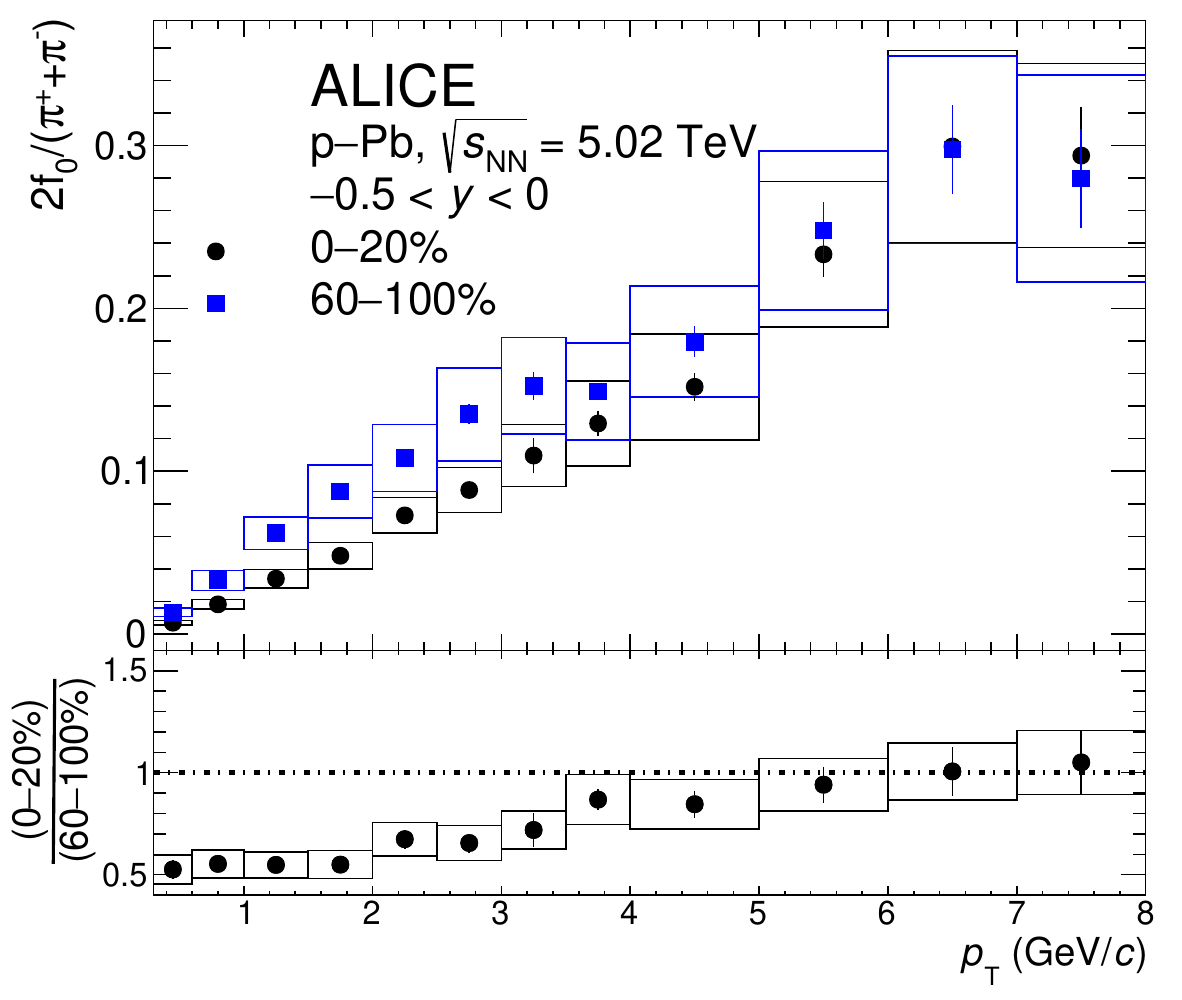} }
	\caption{ The particle yield ratios of \fzero~to $\pi$ as a function of $p_{\rm{T}}$ in high-multiplicity (circles) and low-multiplicity (squares) p--Pb collisions at \snn~=~5.02~TeV. The lower panel shows the double ratio of \fzero/$\pi$ between the high-multiplicity and low-multiplicity events. V0A is utilized to categorize events based on their multiplicity.}
	\label{fig:f0piPt}
\end{figure}

Figure~\ref{fig:f0piPt} shows the $p_{\mathrm{T}}$-differential particle yield ratio of \fzero~to $\pi$ in high-multiplicity (HM, 0--20\%) and low-multiplicity (LM, 60--100\%) p--Pb collisions at \snn~=~5.02~TeV. The ratios are consistent with each other within one sigma at $p_{\mathrm{T}}>$~4~GeV/$c$, while at lower $p_{\mathrm{T}}$ the \fzero~to $\pi$ ratio is systematically lower in the HM class as compared to the LM one. This suppression of \fzero~production at high multiplicity and low $p_{\mathrm{T}}$ is quantified via the double ratio reported in the lower panel of Fig.~\ref{fig:f0piPt}. In the double ratio, the correlated uncertainties across multiplicity classes cancel. The $p_{\mathrm{T}}$ dependence of the double ratio indicates that the suppression of the $p_{\mathrm{T}}$-integrated yield shown in Fig.~\ref{fig:f0piAddCSM} is mainly occurring at low $p_{\mathrm{T}}$ values ($p_{\mathrm{T}}<$~3.5~GeV/$c$), showing a qualitatively similar $p_{\mathrm{T}}$ dependence as the one reported in Ref.~\cite{ALICE:2019etb} for the suppression of the \kstar$\rm{}/K$ ratios.

\begin{figure}[!hbt]
	\centering
	\subfigure{ \includegraphics[width=0.6 \textwidth]{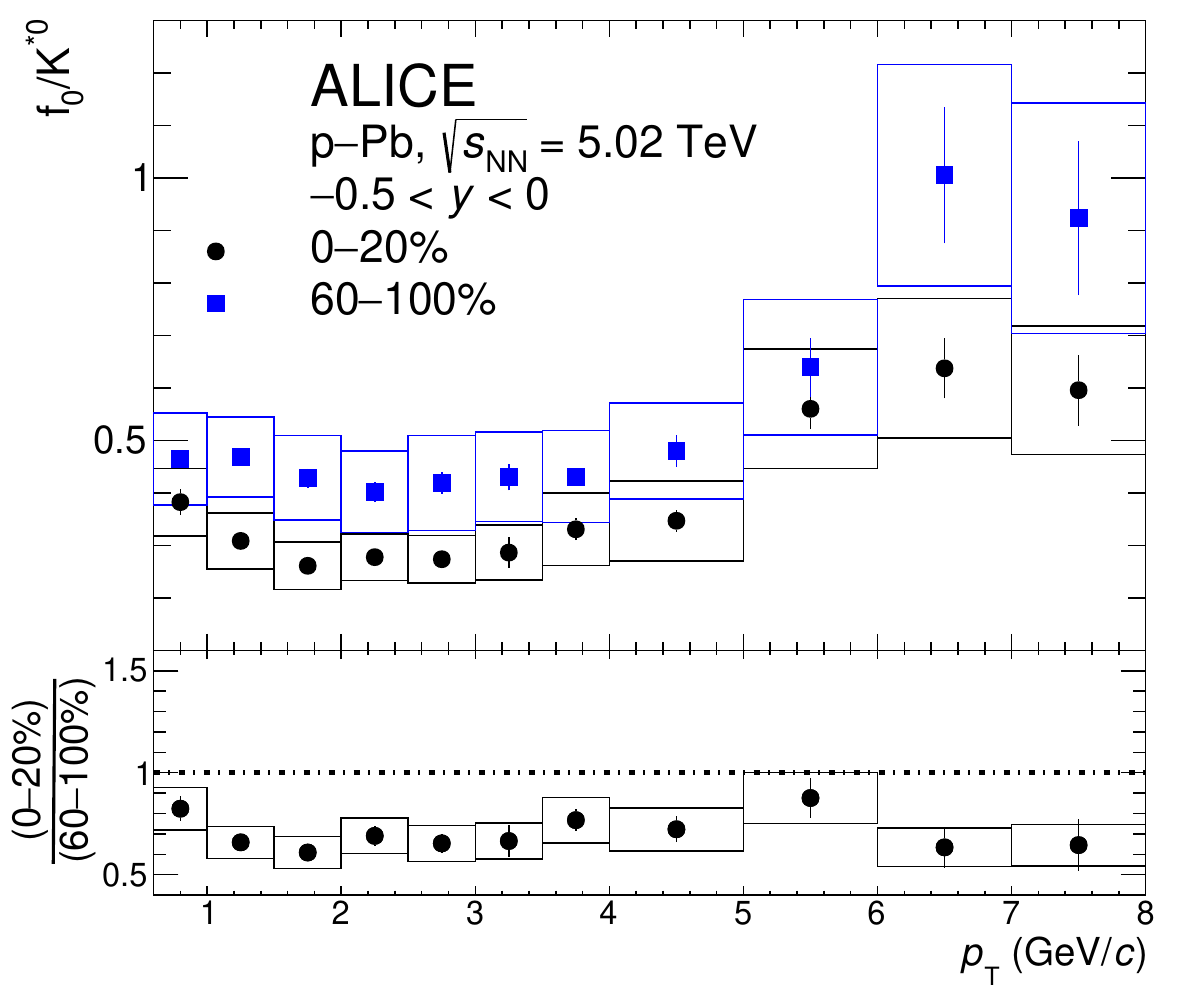} }
	\caption{The particle yield ratio of \fzero~to \kstar~as a function of $p_{\rm{T}}$ in high-multiplicity (circles) and low-multiplicity (squares) p--Pb collisions at \snn~=~5.02~TeV. The lower panel shows the double ratio of high-multiplicity to low-multiplicity \fzero/\kstar. V0A is utilized to categorize events based on their multiplicity. }
	\label{fig:f0KsPt}
\end{figure}

Figure~\ref{fig:f0KsPt} shows the $p_{\mathrm{T}}$-differential particle yield ratio of \fzero~to \kstar~in HM and LM p--Pb collisions at \snn~=~5.02~TeV. The ratio in HM events is lower than that in LM events in the entire $p_{\mathrm{T}}$ range, in contrast to what is observed for \kstar$\rm{}/K$ and \fzero/$\pi$ ratios for which the suppression is observed only at low $p_{\mathrm{T}}$. The suppression of the \fzero/\kstar~ratio in HM events for $p_{\mathrm{T}}>$~4~GeV/$c$ is evaluated to be 0.70~$\pm$~0.04 (stat) $\pm$~0.05 (syst) by fitting the double ratio with a constant function and indicates that other effects, beyond hadronic interactions, are present. For instance, the strangeness enhancement can explain the suppression of the \fzero~yield relative to that of \kstar~under the assumption that the \fzero~does not have strange quark content. This argument is also consistent with the decreasing trend of the $p_{\mathrm{T}}$-integrated \fzero/\kstar~ratio with increasing multiplicity and their comparison to the CSM calculations shown in Fig.~\ref{fig:f0piAddCSM}. In summary, the suppression of the \fzero/\kstar~ratio may suggest that the \fzero~does not contain strange quarks, and its production is therefore not affected by the strangeness enhancement. In addition, the enhancement of baryon-to-meson ratio at intermediate $p_{\mathrm{T}}$, observed for p/$\pi$, $\Lambda$/$\mathrm{K}_{\mathrm{s}}^{0}$, $\Xi$/$\phi$, and $\Omega$/$\phi$ ratios~\cite{ALICE:2020jsh}, is not seen in the \fzero/\kstar~ratio, providing a hint that the number of constituent quarks for \fzero~is two.

\begin{figure}[!hbt]
	\centering
	\subfigure{ \includegraphics[width=0.8 \textwidth]{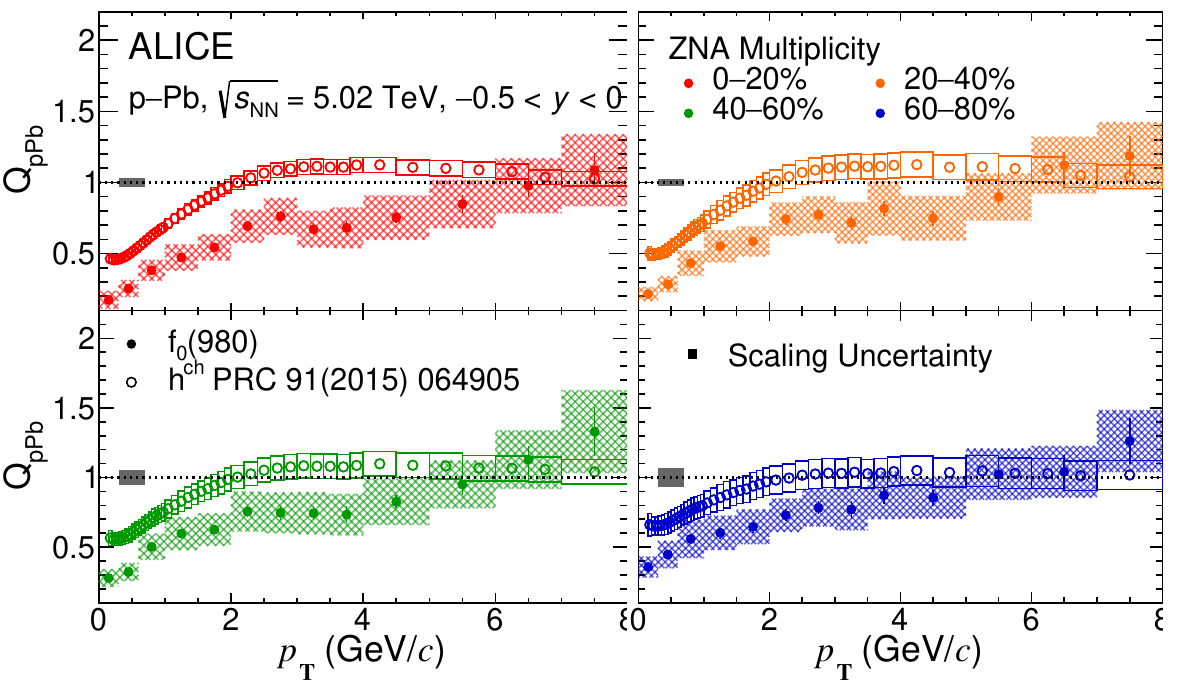} }
	\caption{ Nuclear modification factor ($Q_{\rm{pPb}}$) of \fzero~as a function of $p_{\rm{T}}$ in p--Pb collisions at \snn~=~5.02~TeV for different multiplicity classes. The multiplicity class is defined with the ZNA, which is the ZN placed on the Pb-going side. Statistical and systematic uncertainties are shown as error bars and boxes, respectively. Black boxes around unity represent the binary collision scaling uncertainties. The $Q_{\rm{pPb}}$ of charged hadrons~\cite{ALICE:2014xsp} are reported for comparison. }
	\label{fig:QpPb}
\end{figure}

The $p_{\rm{T}}$-differential yield of \fzero~in p--Pb collisions can be compared to the one in pp collisions at the same center-of-mass energy by computing the nuclear modification factor $Q_{\mathrm{pPb}}$, defined as 
\begin{eqnarray}
Q_{\mathrm{pPb}} = \dfrac{\mathrm{d}^{2} N_{\mathrm{f}_{0}(980)}^{\mathrm{pPb}} / \mathrm{d} p_{\mathrm{T}} \mathrm{d}y }{ \left\langle T_{\mathrm{pPb}} \right\rangle \mathrm{d}^{2} \sigma_{\mathrm{f}_{0}(980)}^{\mathrm{pp}}/ \mathrm{d} p_{\mathrm{T}} \mathrm{d}y },
\end{eqnarray}
where $\left\langle T_{\mathrm{pPb}} \right\rangle$ is the average nuclear overlap function, which is proportional to the number of binary nucleon--nucleon collisions, for the considered centrality class, and $\mathrm{d}^{2} \sigma_{\mathrm{f}_{0}(980)}^{\mathrm{pp}}/ \mathrm{d} p_{\mathrm{T}} \mathrm{d}y$ is the $p_{\mathrm{T}}$ differential cross section for \fzero~production in pp collisions taken from Ref.~\cite{ALICE:2022qnb}. For this study, the centrality classes are defined using an event selection based on the ZN calorimeter in the Pb-going direction (ZNA) to minimize the possible selection biases, as reported in Ref~\cite{ALICE:2014xsp}.

Figure~\ref{fig:QpPb} shows the $Q_{\mathrm{pPb}}$ of \fzero~in p--Pb collisions at \snn~=~5.02~TeV in different multiplicity classes. The systematic uncertainties are calculated with the assumption that there is no correlated uncertainty between the yield in pp and p--Pb collisions except for the B.R. uncertainty, which cancels out in the ratio. The scaling uncertainty on the $Q_{\mathrm{pPb}}$ shown in Fig.~\ref{fig:QpPb} is due to the uncertainty on $\left\langle T_{\mathrm{pPb}} \right\rangle$, which is taken from Ref.~\cite{ALICE:2014xsp}. The $Q_{\mathrm{pPb}}$ distributions of \fzero~are compared to those of charged hadrons~\cite{ALICE:2014xsp}. At low $p_{\mathrm{T}}$ ($p_{\mathrm{T}}<$~4~GeV/$c$), the $Q_{\mathrm{pPb}}$ of \fzero~is lower than unity indicating a suppression of the production in p--Pb collisions relative to pp collisions. This suppression becomes more pronounced with increasing multiplicity for $p_{\mathrm{T}}<$~2~GeV/$c$. Moreover, for $p_{\mathrm{T}}<$~4~GeV/$c$ the $Q_{\mathrm{pPb}}$ of \fzero~is also lower than that of charged hadrons. The \fzero~yield is further suppressed by 30\% in high-multiplicity (0--20\%) events relative to low-multiplicity events (60--80\%), with a 4.7 to 4.9 sigma significance near zero $p_{\mathrm{T}}$. As $p_{\mathrm{T}}$ increases, \fzero~$Q_{\mathrm{pPb}}$ values become compatible with those for charged particles, reaching unity. The dependencies of the nuclear modification factor on the multiplicity and $p_{\mathrm{T}}$ clearly indicate that rescattering largely contributes to the strong suppression of the \fzero~yield for $p_{\mathrm{T}}<$~4~GeV/$c$. In addition, the $Q_{\mathrm{pPb}}$ does not exhibit a significant Cronin-like enhancement~\cite{Cronin:1974zm} at intermediate $p_{\mathrm{T}}$ in HM events. Since baryons show a more pronounced Cronin peak as compared to conventional mesons~\cite{ALICE:2016dei, ALICE:2021rpa}, the absence of a significant Cronin-like enhancement of \fzero~might suggest that the \fzero~is composed of two quarks. 


\section{Conclusions}
\label{sec:summary}

The multiplicity and $p_{\mathrm{T}}$ dependence of \fzero~production in p--Pb collisions at \snn~=~5.02~TeV is presented. The \fzero~is reconstructed via the \fzero~$\rightarrow\pi^{+}\pi^{-}$ decay channel at midrapidity ($-0.5<y<0$) in the transverse momentum region of $0<p_{\mathrm{T}}<8$~GeV/$c$. A hardening of the $p_{\mathrm{T}}$ spectra and a consequent increase of the mean $p_{\mathrm{T}}$ are observed with increasing multiplicity. 

The $p_{\mathrm{T}}$-integrated particle yield ratio of \fzero~to $\pi$ decreases with increasing multiplicity, and the $p_{\mathrm{T}}$-differential studies show a clear suppression of the \fzero~to $\pi$ ratio for $p_{\mathrm{T}}<$~3.5~GeV/$c$, indicating that rescattering effects for \fzero~particles exist in p--Pb collisions. The CSM overestimates the \fzero/$\pi$ ratio, and it does not describe the decreasing trend because the CSM does not consider rescattering processes. The $p_{\mathrm{T}}$-integrated \fzero/\kstar~yield ratio also decreases with increasing multiplicity. The suppression of the \fzero~to \kstar~ratio is observed in the entire measured $p_{\mathrm{T}}$ range, showing a different $p_{\mathrm{T}}$ dependence relative to the one expected from a rescattering scenario. The CSM qualitatively describes the decreasing trend for the $p_{\mathrm{T}}$-integrated \fzero/\kstar~ratio as a function of multiplicity with the assumption of no hidden strangeness for \fzero, while it overestimates the \fzero/\kstar~with the assumption of two strange quarks. These results indicate that the production of \kstar~is relatively enhanced compared with \fzero~due to the strangeness enhancement. 

Additionally, the $p_{\mathrm{T}}$-differential \fzero/\kstar~ratio does not exhibit the characteristic enhancement of baryon-to-meson ratios, suggesting a structure with two constituent quarks for the \fzero~resonance. Furthermore, the multiplicity-dependent nuclear modification factor ($Q_{\mathrm{pPb}}$) for \fzero~exhibits a strong suppression at low $p_{\mathrm{T}}$ with a clear dependence on multiplicity, which can be explained by the rescattering effects. In addition, no Cronin-like enhancement is observed in $Q_{\mathrm{pPb}}$, even in high-multiplicity events. The absence of Cronin-like enhancement in the \fzero~may suggest that the \fzero~is composed of two quarks.

The abnormal suppression in terms of multiplicity and transverse momentum relative to other particles sheds light on the internal structure of \fzero~suggesting that it is a conventional meson with no hidden strange quarks and provides insight into the properties of the late hadronic phase in p--Pb collisions.

\newenvironment{acknowledgement}{\relax}{\relax}
\begin{acknowledgement}
\section*{Acknowledgements}
\noindent 

The ALICE Collaboration would like to thank all its engineers and technicians for their invaluable contributions to the construction of the experiment and the CERN accelerator teams for the outstanding performance of the LHC complex.
The ALICE Collaboration gratefully acknowledges the resources and support provided by all Grid centres and the Worldwide LHC Computing Grid (WLCG) collaboration.
The ALICE Collaboration acknowledges the following funding agencies for their support in building and running the ALICE detector:
A. I. Alikhanyan National Science Laboratory (Yerevan Physics Institute) Foundation (ANSL), State Committee of Science and World Federation of Scientists (WFS), Armenia;
Austrian Academy of Sciences, Austrian Science Fund (FWF): [M 2467-N36] and Nationalstiftung f\"{u}r Forschung, Technologie und Entwicklung, Austria;
Ministry of Communications and High Technologies, National Nuclear Research Center, Azerbaijan;
Conselho Nacional de Desenvolvimento Cient\'{\i}fico e Tecnol\'{o}gico (CNPq), Financiadora de Estudos e Projetos (Finep), Funda\c{c}\~{a}o de Amparo \`{a} Pesquisa do Estado de S\~{a}o Paulo (FAPESP) and Universidade Federal do Rio Grande do Sul (UFRGS), Brazil;
Bulgarian Ministry of Education and Science, within the National Roadmap for Research Infrastructures 2020-2027 (object CERN), Bulgaria;
Ministry of Education of China (MOEC) , Ministry of Science \& Technology of China (MSTC) and National Natural Science Foundation of China (NSFC), China;
Ministry of Science and Education and Croatian Science Foundation, Croatia;
Centro de Aplicaciones Tecnol\'{o}gicas y Desarrollo Nuclear (CEADEN), Cubaenerg\'{\i}a, Cuba;
Ministry of Education, Youth and Sports of the Czech Republic, Czech Republic;
The Danish Council for Independent Research | Natural Sciences, the VILLUM FONDEN and Danish National Research Foundation (DNRF), Denmark;
Helsinki Institute of Physics (HIP), Finland;
Commissariat \`{a} l'Energie Atomique (CEA) and Institut National de Physique Nucl\'{e}aire et de Physique des Particules (IN2P3) and Centre National de la Recherche Scientifique (CNRS), France;
Bundesministerium f\"{u}r Bildung und Forschung (BMBF) and GSI Helmholtzzentrum f\"{u}r Schwerionenforschung GmbH, Germany;
General Secretariat for Research and Technology, Ministry of Education, Research and Religions, Greece;
National Research, Development and Innovation Office, Hungary;
Department of Atomic Energy Government of India (DAE), Department of Science and Technology, Government of India (DST), University Grants Commission, Government of India (UGC) and Council of Scientific and Industrial Research (CSIR), India;
National Research and Innovation Agency - BRIN, Indonesia;
Istituto Nazionale di Fisica Nucleare (INFN), Italy;
Japanese Ministry of Education, Culture, Sports, Science and Technology (MEXT) and Japan Society for the Promotion of Science (JSPS) KAKENHI, Japan;
Consejo Nacional de Ciencia (CONACYT) y Tecnolog\'{i}a, through Fondo de Cooperaci\'{o}n Internacional en Ciencia y Tecnolog\'{i}a (FONCICYT) and Direcci\'{o}n General de Asuntos del Personal Academico (DGAPA), Mexico;
Nederlandse Organisatie voor Wetenschappelijk Onderzoek (NWO), Netherlands;
The Research Council of Norway, Norway;
Commission on Science and Technology for Sustainable Development in the South (COMSATS), Pakistan;
Pontificia Universidad Cat\'{o}lica del Per\'{u}, Peru;
Ministry of Education and Science, National Science Centre and WUT ID-UB, Poland;
Korea Institute of Science and Technology Information and National Research Foundation of Korea (NRF), Republic of Korea;
Ministry of Education and Scientific Research, Institute of Atomic Physics, Ministry of Research and Innovation and Institute of Atomic Physics and Universitatea Nationala de Stiinta si Tehnologie Politehnica Bucuresti, Romania;
Ministry of Education, Science, Research and Sport of the Slovak Republic, Slovakia;
National Research Foundation of South Africa, South Africa;
Swedish Research Council (VR) and Knut \& Alice Wallenberg Foundation (KAW), Sweden;
European Organization for Nuclear Research, Switzerland;
Suranaree University of Technology (SUT), National Science and Technology Development Agency (NSTDA) and National Science, Research and Innovation Fund (NSRF via PMU-B B05F650021), Thailand;
Turkish Energy, Nuclear and Mineral Research Agency (TENMAK), Turkey;
National Academy of  Sciences of Ukraine, Ukraine;
Science and Technology Facilities Council (STFC), United Kingdom;
National Science Foundation of the United States of America (NSF) and United States Department of Energy, Office of Nuclear Physics (DOE NP), United States of America.
In addition, individual groups or members have received support from:
Czech Science Foundation (grant no. 23-07499S), Czech Republic;
European Research Council, Strong 2020 - Horizon 2020 (grant nos. 950692, 824093), European Union;
ICSC - Centro Nazionale di Ricerca in High Performance Computing, Big Data and Quantum Computing, European Union - NextGenerationEU;
Academy of Finland (Center of Excellence in Quark Matter) (grant nos. 346327, 346328), Finland.
\end{acknowledgement}

\bibliographystyle{utphys}
\bibliography{paper.bib}

\providecommand{\href}[2]{#2}\begingroup\raggedright\begin{thebibliography}{10}

\bibitem{ParticleDataGroup:2022pth}
{\bfseries Particle Data Group} Collaboration, R.~L. Workman {\em et~al.},
  ``{Review of Particle Physics}'',
  \href{http://dx.doi.org/10.1093/ptep/ptac097}{{\em PTEP} {\bfseries 2022}
  (2022) 083C01}.

\bibitem{ExHIC:2010gcb}
{\bfseries ExHIC} Collaboration, S.~Cho {\em et~al.}, ``{Multi-quark hadrons
  from Heavy Ion Collisions}'',
  \href{http://dx.doi.org/10.1103/PhysRevLett.106.212001}{{\em Phys. Rev.
  Lett.} {\bfseries 106} (2011) 212001},
  \href{http://arxiv.org/abs/1011.0852}{{\ttfamily arXiv:1011.0852 [nucl-th]}}.

\bibitem{Jaffe:1976ig}
R.~L. Jaffe, ``{Multi-Quark Hadrons. 1. The Phenomenology of (2 Quark 2
  anti-Quark) Mesons}'', \href{http://dx.doi.org/10.1103/PhysRevD.15.267}{{\em
  Phys. Rev. D} {\bfseries 15} (1977) 267}.

\bibitem{Maiani:2004uc}
L.~Maiani, F.~Piccinini, A.~D. Polosa, and V.~Riquer, ``{A New look at scalar
  mesons}'', \href{http://dx.doi.org/10.1103/PhysRevLett.93.212002}{{\em Phys.
  Rev. Lett.} {\bfseries 93} (2004) 212002},
  \href{http://arxiv.org/abs/hep-ph/0407017}{{\ttfamily arXiv:hep-ph/0407017}}.

\bibitem{Chen:2003za}
C.-H. Chen, ``{Evidence for two quark content of f(0)(980) in exclusive b
  ---\ensuremath{>} c decays}'',
  \href{http://dx.doi.org/10.1103/PhysRevD.67.094011}{{\em Phys. Rev. D}
  {\bfseries 67} (2003) 094011},
  \href{http://arxiv.org/abs/hep-ph/0302059}{{\ttfamily arXiv:hep-ph/0302059}}.

\bibitem{Achasov:2020aun}
N.~N. Achasov, J.~V. Bennett, A.~V. Kiselev, E.~A. Kozyrev, and G.~N.
  Shestakov, ``{Evidence of the four-quark nature of $f_0$(980) and
  $f_0$(500)}'', \href{http://dx.doi.org/10.1103/PhysRevD.103.014010}{{\em
  Phys. Rev. D} {\bfseries 103} (2021) 014010},
  \href{http://arxiv.org/abs/2009.04191}{{\ttfamily arXiv:2009.04191
  [hep-ph]}}.

\bibitem{Ahmed:2020kmp}
H.~A. Ahmed and C.~W. Xiao, ``{Study the molecular nature of $\sigma$,
  $f_{0}(980)$, and $a_{0}(980)$ states}'',
  \href{http://dx.doi.org/10.1103/PhysRevD.101.094034}{{\em Phys. Rev. D}
  {\bfseries 101} (2020) 094034},
  \href{http://arxiv.org/abs/2001.08141}{{\ttfamily arXiv:2001.08141
  [hep-ph]}}.

\bibitem{Bhalerao:2020ulk}
R.~S. Bhalerao, ``{Collectivity in large and small systems formed in
  ultrarelativistic collisions}'',
  \href{http://dx.doi.org/10.1140/epjs/s11734-021-00019-x}{{\em Eur. Phys. J.
  ST} {\bfseries 230} (2021) 635--654},
  \href{http://arxiv.org/abs/2009.09586}{{\ttfamily arXiv:2009.09586
  [nucl-th]}}.

\bibitem{ALICE:2019zfl}
{\bfseries ALICE} Collaboration, S.~Acharya {\em et~al.}, ``{Investigations of
  Anisotropic Flow Using Multiparticle Azimuthal Correlations in pp, p--Pb,
  Xe--Xe, and Pb--Pb Collisions at the LHC}'',
  \href{http://dx.doi.org/10.1103/PhysRevLett.123.142301}{{\em Phys. Rev.
  Lett.} {\bfseries 123} (2019) 142301},
  \href{http://arxiv.org/abs/1903.01790}{{\ttfamily arXiv:1903.01790
  [nucl-ex]}}.

\bibitem{Adams:2005dq}
{\bfseries STAR} Collaboration, J.~Adams {\em et~al.}, ``{Experimental and
  theoretical challenges in the search for the quark gluon plasma: The STAR
  Collaboration's critical assessment of the evidence from RHIC collisions}'',
  \href{http://dx.doi.org/10.1016/j.nuclphysa.2005.03.085}{{\em Nucl. Phys.}
  {\bfseries A757} (2005) 102--183},
\href{http://arxiv.org/abs/nucl-ex/0501009}{{\ttfamily arXiv:nucl-ex/0501009
  [nucl-ex]}}.

\bibitem{Adcox:2004mh}
{\bfseries PHENIX} Collaboration, K.~Adcox {\em et~al.}, ``{Formation of dense
  partonic matter in relativistic nucleus-nucleus collisions at RHIC:
  Experimental evaluation by the PHENIX collaboration}'',
  \href{http://dx.doi.org/10.1016/j.nuclphysa.2005.03.086}{{\em Nucl. Phys.}
  {\bfseries A757} (2005) 184--283},
\href{http://arxiv.org/abs/nucl-ex/0410003}{{\ttfamily arXiv:nucl-ex/0410003
  [nucl-ex]}}.

\bibitem{ALICE:2019qyj}
{\bfseries ALICE} Collaboration, S.~Acharya {\em et~al.}, ``{Measurements of
  inclusive jet spectra in pp and central Pb--Pb collisions at
  $\sqrt{s_{\rm{NN}}}$ = 5.02 TeV}'',
  \href{http://dx.doi.org/10.1103/PhysRevC.101.034911}{{\em Phys. Rev. C}
  {\bfseries 101} (2020) 034911},
  \href{http://arxiv.org/abs/1909.09718}{{\ttfamily arXiv:1909.09718
  [nucl-ex]}}.

\bibitem{ATLAS:2010isq}
{\bfseries ATLAS} Collaboration, G.~Aad {\em et~al.}, ``{Observation of a
  Centrality-Dependent Dijet Asymmetry in Lead-Lead Collisions at
  $\sqrt{s_{\mathrm{NN}}}=2.77$ TeV with the ATLAS Detector at the LHC}'',
  \href{http://dx.doi.org/10.1103/PhysRevLett.105.252303}{{\em Phys. Rev.
  Lett.} {\bfseries 105} (2010) 252303},
  \href{http://arxiv.org/abs/1011.6182}{{\ttfamily arXiv:1011.6182 [hep-ex]}}.

\bibitem{PHENIX:2010nlr}
{\bfseries PHENIX} Collaboration, A.~Adare {\em et~al.}, ``{Azimuthal
  anisotropy of neutral pion production in Au+Au collisions at
  $\sqrt{s_{\mathrm{NN}}}$ = 200 GeV: Path-length dependence of jet quenching
  and the role of initial geometry}'',
  \href{http://dx.doi.org/10.1103/PhysRevLett.105.142301}{{\em Phys. Rev.
  Lett.} {\bfseries 105} (2010) 142301},
  \href{http://arxiv.org/abs/1006.3740}{{\ttfamily arXiv:1006.3740 [nucl-ex]}}.

\bibitem{ALICE:2019hno}
{\bfseries ALICE} Collaboration, S.~Acharya {\em et~al.}, ``{Production of
  charged pions, kaons, and (anti-)protons in Pb--Pb and inelastic pp
  collisions at $\sqrt{s_{\mathrm{NN}}}$ = 5.02 TeV}'',
  \href{http://dx.doi.org/10.1103/PhysRevC.101.044907}{{\em Phys. Rev. C}
  {\bfseries 101} (2020) 044907},
  \href{http://arxiv.org/abs/1910.07678}{{\ttfamily arXiv:1910.07678
  [nucl-ex]}}.

\bibitem{PHENIX:2006ujp}
{\bfseries PHENIX} Collaboration, S.~S. Adler {\em et~al.}, ``{Common
  suppression pattern of $\eta$ and $\pi^{0}$ mesons at high transverse
  momentum in Au+Au collisions at $\sqrt{s_{\mathrm{NN}}}=$~200~GeV}'',
  \href{http://dx.doi.org/10.1103/PhysRevLett.96.202301}{{\em Phys. Rev. Lett.}
  {\bfseries 96} (2006) 202301},
  \href{http://arxiv.org/abs/nucl-ex/0601037}{{\ttfamily
  arXiv:nucl-ex/0601037}}.

\bibitem{Heinz:2000bk}
U.~W. Heinz and M.~Jacob, ``{Evidence for a new state of matter: An Assessment
  of the results from the CERN lead beam program}'',
  \href{http://arxiv.org/abs/nucl-th/0002042}{{\ttfamily
  arXiv:nucl-th/0002042}}.

\bibitem{ALICE:2022wpn}
{\bfseries ALICE} Collaboration, ``{The ALICE experiment -- A journey through
  QCD}'', \href{http://arxiv.org/abs/2211.04384}{{\ttfamily arXiv:2211.04384
  [nucl-ex]}}.

\bibitem{ALICE:2016dei}
{\bfseries ALICE} Collaboration, J.~Adam {\em et~al.}, ``{Multiplicity
  dependence of charged pion, kaon, and (anti)proton production at large
  transverse momentum in p--Pb collisions at $\mathbf{\sqrt{{\textit s}_{\rm
  NN}}}$ = 5.02 TeV}'',
  \href{http://dx.doi.org/10.1016/j.physletb.2016.07.050}{{\em Phys. Lett. B}
  {\bfseries 760} (2016) 720--735},
  \href{http://arxiv.org/abs/1601.03658}{{\ttfamily arXiv:1601.03658
  [nucl-ex]}}.

\bibitem{ALICE:2016fzo}
{\bfseries ALICE} Collaboration, J.~Adam {\em et~al.}, ``{Enhanced production
  of multi-strange hadrons in high-multiplicity proton-proton collisions}'',
  \href{http://dx.doi.org/10.1038/nphys4111}{{\em Nature Phys.} {\bfseries 13}
  (2017) 535--539}, \href{http://arxiv.org/abs/1606.07424}{{\ttfamily
  arXiv:1606.07424 [nucl-ex]}}.

\bibitem{LHCb:2014ooi}
{\bfseries LHCb} Collaboration, R.~Aaij {\em et~al.}, ``{Measurement of
  resonant and CP components in $\bar{B}_s^0\to J/\psi\pi^+\pi^-$ decays}'',
  \href{http://dx.doi.org/10.1103/PhysRevD.89.092006}{{\em Phys. Rev. D}
  {\bfseries 89} (2014) 092006},
  \href{http://arxiv.org/abs/1402.6248}{{\ttfamily arXiv:1402.6248 [hep-ex]}}.

\bibitem{LHCb:2014vbo}
{\bfseries LHCb} Collaboration, R.~Aaij {\em et~al.}, ``{Measurement of the
  resonant and CP components in $\overline{B}^0\to J/\psi \pi^+\pi^-$
  decays}'', \href{http://dx.doi.org/10.1103/PhysRevD.90.012003}{{\em Phys.
  Rev. D} {\bfseries 90} (2014) 012003},
  \href{http://arxiv.org/abs/1404.5673}{{\ttfamily arXiv:1404.5673 [hep-ex]}}.

\bibitem{Fries:2003vb}
R.~J. Fries, B.~Muller, C.~Nonaka, and S.~A. Bass, ``{Hadronization in heavy
  ion collisions: Recombination and fragmentation of partons}'',
  \href{http://dx.doi.org/10.1103/PhysRevLett.90.202303}{{\em Phys. Rev. Lett.}
  {\bfseries 90} (2003) 202303},
  \href{http://arxiv.org/abs/nucl-th/0301087}{{\ttfamily
  arXiv:nucl-th/0301087}}.

\bibitem{Wang:2022det}
M.~Wang, J.-Q. Tao, H.~Zheng, W.-C. Zhang, L.-L. Zhu, and A.~Bonasera,
  ``{Number-of-constituent-quark scaling of elliptic flow: a quantitative
  study}'', \href{http://dx.doi.org/10.1007/s41365-022-01019-9}{{\em Nucl. Sci.
  Tech.} {\bfseries 33} (2022) 37},
  \href{http://arxiv.org/abs/2203.10353}{{\ttfamily arXiv:2203.10353
  [hep-ph]}}.

\bibitem{ALICE:2018qdv}
{\bfseries ALICE} Collaboration, S.~Acharya {\em et~al.}, ``{Production of the
  $\rho$(770)${^{0}}$ meson in pp and Pb--Pb collisions at $\sqrt{s_{\rm NN}}$
  = 2.76 TeV}'', \href{http://dx.doi.org/10.1103/PhysRevC.99.064901}{{\em Phys.
  Rev. C} {\bfseries 99} (2019) 064901},
  \href{http://arxiv.org/abs/1805.04365}{{\ttfamily arXiv:1805.04365
  [nucl-ex]}}.

\bibitem{ALICE:2019etb}
{\bfseries ALICE} Collaboration, S.~Acharya {\em et~al.}, ``{Multiplicity
  dependence of K*(892)$^{0}$ and $\phi$(1020) production in pp collisions at
  $\sqrt {s}$ =~13 TeV}'',
  \href{http://dx.doi.org/10.1016/j.physletb.2020.135501}{{\em Phys. Lett. B}
  {\bfseries 807} (2020) 135501},
  \href{http://arxiv.org/abs/1910.14397}{{\ttfamily arXiv:1910.14397
  [nucl-ex]}}.

\bibitem{ALICE:2016sak}
{\bfseries ALICE} Collaboration, J.~Adam {\em et~al.}, ``{Production of
  K$^{*}$(892)$^{0}$ and $\phi$(1020) in p\textendash{}Pb collisions at
  $\sqrt{s_{{\text {NN}}}}$ = 5.02 TeV}'',
  \href{http://dx.doi.org/10.1140/epjc/s10052-016-4088-7}{{\em Eur. Phys. J. C}
  {\bfseries 76} (2016) 245}, \href{http://arxiv.org/abs/1601.07868}{{\ttfamily
  arXiv:1601.07868 [nucl-ex]}}.

\bibitem{ALICE:2022zuc}
{\bfseries ALICE} Collaboration, S.~Acharya {\em et~al.}, ``{$\Sigma
  (1385)^{\pm }$ resonance production in Pb\textendash{}Pb collisions at
  $\sqrt{s_{\textrm{NN}}}~=~5.02$~TeV}'',
  \href{http://dx.doi.org/10.1140/epjc/s10052-023-11475-1}{{\em Eur. Phys. J.
  C} {\bfseries 83} (2023) 351},
  \href{http://arxiv.org/abs/2205.13998}{{\ttfamily arXiv:2205.13998
  [nucl-ex]}}.

\bibitem{ALICE:2018ewo}
{\bfseries ALICE} Collaboration, S.~Acharya {\em et~al.}, ``{Suppression of
  $\Lambda(1520)$ resonance production in central Pb--Pb collisions at
  $\sqrt{s_{\rm NN}}$ = 2.76 TeV}'',
  \href{http://dx.doi.org/10.1103/PhysRevC.99.024905}{{\em Phys. Rev. C}
  {\bfseries 99} (2019) 024905},
  \href{http://arxiv.org/abs/1805.04361}{{\ttfamily arXiv:1805.04361
  [nucl-ex]}}.

\bibitem{Bierlich:2021poz}
C.~Bierlich, T.~Sj\"ostrand, and M.~Utheim, ``{Hadronic rescattering in pA and
  AA collisions}'',
  \href{http://dx.doi.org/10.1140/epja/s10050-021-00543-3}{{\em Eur. Phys. J.
  A} {\bfseries 57} (2021) 227},
  \href{http://arxiv.org/abs/2103.09665}{{\ttfamily arXiv:2103.09665
  [hep-ph]}}.

\bibitem{Knospe:2015nva}
A.~G. Knospe, C.~Markert, K.~Werner, J.~Steinheimer, and M.~Bleicher,
  ``{Hadronic resonance production and interaction in partonic and hadronic
  matter in the EPOS3 model with and without the hadronic afterburner UrQMD}'',
  \href{http://dx.doi.org/10.1103/PhysRevC.93.014911}{{\em Phys. Rev. C}
  {\bfseries 93} (2016) 014911},
  \href{http://arxiv.org/abs/1509.07895}{{\ttfamily arXiv:1509.07895
  [nucl-th]}}.

\bibitem{Song:1996ik}
C.~Song and V.~Koch, ``{Chemical relaxation time of pions in hot hadronic
  matter}'', \href{http://dx.doi.org/10.1103/PhysRevC.55.3026}{{\em Phys. Rev.
  C} {\bfseries 55} (1997) 3026--3037},
  \href{http://arxiv.org/abs/nucl-th/9611034}{{\ttfamily
  arXiv:nucl-th/9611034}}.

\bibitem{ALICE:2011dyt}
{\bfseries ALICE} Collaboration, K.~Aamodt {\em et~al.}, ``{Two-pion
  Bose-Einstein correlations in central Pb--Pb collisions at
  $\sqrt{s_{\mathrm{NN}}} =$ 2.76 TeV}'',
  \href{http://dx.doi.org/10.1016/j.physletb.2010.12.053}{{\em Phys. Lett. B}
  {\bfseries 696} (2011) 328--337},
  \href{http://arxiv.org/abs/1012.4035}{{\ttfamily arXiv:1012.4035 [nucl-ex]}}.

\bibitem{ALICE:2019xyr}
{\bfseries ALICE} Collaboration, S.~Acharya {\em et~al.}, ``{Evidence of
  rescattering effect in Pb--Pb collisions at the LHC through production of
  $\rm{K}^{*}(892)^{0}$ and $\phi(1020)$ mesons}'',
  \href{http://dx.doi.org/10.1016/j.physletb.2020.135225}{{\em Phys. Lett. B}
  {\bfseries 802} (2020) 135225},
  \href{http://arxiv.org/abs/1910.14419}{{\ttfamily arXiv:1910.14419
  [nucl-ex]}}.

\bibitem{ALICE:2018pal}
{\bfseries ALICE} Collaboration, S.~Acharya {\em et~al.}, ``{Multiplicity
  dependence of light-flavor hadron production in pp collisions at $\sqrt{s}$ =
  7 TeV}'', \href{http://dx.doi.org/10.1103/PhysRevC.99.024906}{{\em Phys. Rev.
  C} {\bfseries 99} (2019) 024906},
  \href{http://arxiv.org/abs/1807.11321}{{\ttfamily arXiv:1807.11321
  [nucl-ex]}}.

\bibitem{Abelev:2014ffa}
{\bfseries ALICE} Collaboration, B.~B. Abelev {\em et~al.}, ``{Performance of
  the ALICE Experiment at the CERN LHC}'',
  \href{http://dx.doi.org/10.1142/S0217751X14300440}{{\em Int. J. Mod. Phys.}
  {\bfseries A29} (2014) 1430044},
\href{http://arxiv.org/abs/1402.4476}{{\ttfamily arXiv:1402.4476 [nucl-ex]}}.

\bibitem{ALICE:2013axi}
{\bfseries ALICE} Collaboration, E.~Abbas {\em et~al.}, ``{Performance of the
  ALICE VZERO system}'',
  \href{http://dx.doi.org/10.1088/1748-0221/8/10/P10016}{{\em JINST} {\bfseries
  8} (2013) P10016}, \href{http://arxiv.org/abs/1306.3130}{{\ttfamily
  arXiv:1306.3130 [nucl-ex]}}.

\bibitem{Cortese:2019nnv}
{\bfseries ALICE} Collaboration, P.~Cortese, ``{Performance of the ALICE Zero
  Degree Calorimeters and upgrade strategy}'',
  \href{http://dx.doi.org/10.1088/1742-6596/1162/1/012006}{{\em J. Phys. Conf.
  Ser.} {\bfseries 1162} (2019) 012006}.

\bibitem{ALICE:2010tia}
{\bfseries ALICE} Collaboration, K.~Aamodt {\em et~al.}, ``{Alignment of the
  ALICE Inner Tracking System with cosmic-ray tracks}'',
  \href{http://dx.doi.org/10.1088/1748-0221/5/03/P03003}{{\em JINST} {\bfseries
  5} (2010) P03003}, \href{http://arxiv.org/abs/1001.0502}{{\ttfamily
  arXiv:1001.0502 [physics.ins-det]}}.

\bibitem{Alme:2010ke}
J.~Alme {\em et~al.}, ``{The ALICE TPC, a large 3-dimensional tracking device
  with fast readout for ultra-high multiplicity events}'',
  \href{http://dx.doi.org/10.1016/j.nima.2010.04.042}{{\em Nucl. Instrum. Meth.
  A} {\bfseries 622} (2010) 316--367},
  \href{http://arxiv.org/abs/1001.1950}{{\ttfamily arXiv:1001.1950
  [physics.ins-det]}}.

\bibitem{Jacazio:2018slq}
{\bfseries ALICE} Collaboration, N.~Jacazio, ``{PID performance of the
  ALICE-TOF detector at Run 2}'',
  \href{http://dx.doi.org/10.22323/1.321.0232}{{\em PoS} {\bfseries LHCP2018}
  (2018) 232}, \href{http://arxiv.org/abs/1809.00574}{{\ttfamily
  arXiv:1809.00574 [physics.ins-det]}}.

\bibitem{ALICE:2014gvw}
{\bfseries ALICE} Collaboration, B.~B. Abelev {\em et~al.}, ``{Measurement of
  visible cross sections in proton-lead collisions at $\sqrt{s_{\rm NN}}$ =
  5.02 TeV in van der Meer scans with the ALICE detector}'',
  \href{http://dx.doi.org/10.1088/1748-0221/9/11/P11003}{{\em JINST} {\bfseries
  9} (2014) P11003}, \href{http://arxiv.org/abs/1405.1849}{{\ttfamily
  arXiv:1405.1849 [nucl-ex]}}.

\bibitem{ALICE:2014xsp}
{\bfseries ALICE} Collaboration, J.~Adam {\em et~al.}, ``{Centrality dependence
  of particle production in p--Pb collisions at $\sqrt{s_{\rm NN} }$= 5.02
  TeV}'', \href{http://dx.doi.org/10.1103/PhysRevC.91.064905}{{\em Phys. Rev.
  C} {\bfseries 91} (2015) 064905},
  \href{http://arxiv.org/abs/1412.6828}{{\ttfamily arXiv:1412.6828 [nucl-ex]}}.

\bibitem{Santoro2009:ALICESPD}
R.~Santoro {\em et~al.}, ``{The ALICE Silicon Pixel Detector: Readiness for the
  first proton beam}'',
  \href{http://dx.doi.org/10.1088/1748-0221/4/03/P03023}{{\em JINST} {\bfseries
  4} (2009) P03023}.

\bibitem{ALICE:2015olq}
{\bfseries ALICE} Collaboration, J.~Adam {\em et~al.}, ``{Charged-particle
  multiplicities in proton--proton collisions at $\sqrt{s} = 0.9$ to 8 TeV}'',
  \href{http://dx.doi.org/10.1140/epjc/s10052-016-4571-1}{{\em Eur. Phys. J. C}
  {\bfseries 77} (2017) 33}, \href{http://arxiv.org/abs/1509.07541}{{\ttfamily
  arXiv:1509.07541 [nucl-ex]}}.

\bibitem{ALICE:2017svf}
{\bfseries ALICE} Collaboration, S.~Acharya {\em et~al.}, ``{Constraints on jet
  quenching in p--Pb collisions at $\sqrt{s_{\mathrm{NN}}}$ = 5.02 TeV measured
  by the event-activity dependence of semi-inclusive hadron-jet
  distributions}'',
  \href{http://dx.doi.org/10.1016/j.physletb.2018.05.059}{{\em Phys. Lett. B}
  {\bfseries 783} (2018) 95--113},
  \href{http://arxiv.org/abs/1712.05603}{{\ttfamily arXiv:1712.05603
  [nucl-ex]}}.

\bibitem{Stone:2013eaa}
S.~Stone and L.~Zhang, ``{Use of $B\to J/\psi f_0$ decays to discern the $q
  \bar{q}$ or tetraquark nature of scalar mesons}'',
  \href{http://dx.doi.org/10.1103/PhysRevLett.111.062001}{{\em Phys. Rev.
  Lett.} {\bfseries 111} (2013) 062001},
  \href{http://arxiv.org/abs/1305.6554}{{\ttfamily arXiv:1305.6554 [hep-ex]}}.

\bibitem{ALICE:2022qnb}
{\bfseries ALICE} Collaboration, S.~Acharya {\em et~al.}, ``{${\rm f}_{0}(980)$
  production in inelastic pp collisions at $\sqrt{s} = 5.02$ TeV}'',
  \href{http://dx.doi.org/10.1016/j.physletb.2022.137644}{{\em Phys. Lett. B}
  {\bfseries 846} (2023) 137644},
  \href{http://arxiv.org/abs/2206.06216}{{\ttfamily arXiv:2206.06216
  [nucl-ex]}}.

\bibitem{ALICE:2013wgn}
{\bfseries ALICE} Collaboration, B.~B. Abelev {\em et~al.}, ``{Multiplicity
  Dependence of Pion, Kaon, Proton and Lambda Production in p--Pb Collisions at
  $\sqrt{s_{\mathrm{NN}}}$ = 5.02 TeV}'',
  \href{http://dx.doi.org/10.1016/j.physletb.2013.11.020}{{\em Phys. Lett. B}
  {\bfseries 728} (2014) 25--38},
  \href{http://arxiv.org/abs/1307.6796}{{\ttfamily arXiv:1307.6796 [nucl-ex]}}.

\bibitem{LIKESIGN}
T.~Sj\"ostrand and M.~van Zijl, ``A multiple-interaction model for the event
  structure in hadron collisions'',
  \href{http://dx.doi.org/10.1103/PhysRevD.36.2019}{{\em Phys. Rev. D}
  {\bfseries 36} (Oct, 1987) 2019--2041}.

\bibitem{OPAL:1998enc}
{\bfseries OPAL} Collaboration, K.~Ackerstaff {\em et~al.}, ``{Photon and light
  meson production in hadronic Z$^{0}$ decays}'',
  \href{http://dx.doi.org/10.1007/s100520050286}{{\em Eur. Phys. J. C}
  {\bfseries 5} (1998) 411--437},
  \href{http://arxiv.org/abs/hep-ex/9805011}{{\ttfamily arXiv:hep-ex/9805011}}.

\bibitem{STAR:2003vqj}
{\bfseries STAR} Collaboration, J.~Adams {\em et~al.}, ``{$\rho^{0}$ production
  and possible modification in Au+Au and p+p collisions at
  $\sqrt{s_{\mathrm{NN}}}$ = 200~GeV}'',
  \href{http://dx.doi.org/10.1103/PhysRevLett.92.092301}{{\em Phys. Rev. Lett.}
  {\bfseries 92} (2004) 092301},
  \href{http://arxiv.org/abs/nucl-ex/0307023}{{\ttfamily
  arXiv:nucl-ex/0307023}}.

\bibitem{Fedynitch:2015kcn}
A.~Fedynitch, \href{http://dx.doi.org/10.5445/IR/1000055433}{{\em {Cascade
  equations and hadronic interactions at very high energies}}}.
\newblock PhD thesis, KIT, Karlsruhe, Dept. Phys., 11, CERN-THESIS-2015-371.

\bibitem{Brun:1994aa}
R.~Brun, F.~Bruyant, F.~Carminati, S.~Giani, M.~Maire, A.~McPherson,
  G.~Patrick, and L.~Urban, ``{GEANT Detector Description and Simulation
  Tool}'', \href{http://dx.doi.org/10.17181/CERN.MUHF.DMJ1}{{\em CERN-W5013}
  (10, 1994) }.

\bibitem{Altenkamper:2017qot}
L.~Altenk\"amper, F.~Bock, C.~Loizides, and N.~Schmidt, ``{Applicability of
  transverse mass scaling in hadronic collisions at energies available at the
  CERN Large Hadron Collider}'',
  \href{http://dx.doi.org/10.1103/PhysRevC.96.064907}{{\em Phys. Rev. C}
  {\bfseries 96} (2017) 064907},
  \href{http://arxiv.org/abs/1710.01933}{{\ttfamily arXiv:1710.01933
  [hep-ph]}}.

\bibitem{Schnedermann:1993ws}
E.~Schnedermann, J.~Sollfrank, and U.~W. Heinz, ``{Thermal phenomenology of
  hadrons from 200-A/GeV S+S collisions}'',
  \href{http://dx.doi.org/10.1103/PhysRevC.48.2462}{{\em Phys. Rev. C}
  {\bfseries 48} (1993) 2462--2475},
  \href{http://arxiv.org/abs/nucl-th/9307020}{{\ttfamily
  arXiv:nucl-th/9307020}}.

\bibitem{ALICE:2012xs}
{\bfseries ALICE} Collaboration, B.~Abelev {\em et~al.}, ``{Pseudorapidity
  density of charged particles in p--Pb collisions at
  $\sqrt{s_{\mathrm{NN}}}=$~5.02~TeV}'',
  \href{http://dx.doi.org/10.1103/PhysRevLett.110.032301}{{\em Phys. Rev.
  Lett.} {\bfseries 110} (2013) 032301},
  \href{http://arxiv.org/abs/1210.3615}{{\ttfamily arXiv:1210.3615 [nucl-ex]}}.

\bibitem{Liu:2018xae}
P.~Liu and R.~A. Lacey, ``{System-size dependence of the viscous attenuation of
  anisotropic flow in $p$ + Pb and Pb + Pb collisions at energies available at
  the CERN Large Hadron Collider}'',
  \href{http://dx.doi.org/10.1103/PhysRevC.98.031901}{{\em Phys. Rev. C}
  {\bfseries 98} (2018) 031901},
  \href{http://arxiv.org/abs/1804.04618}{{\ttfamily arXiv:1804.04618
  [nucl-ex]}}.

\bibitem{Vovchenko:2019kes}
V.~Vovchenko, B.~D\"onigus, and H.~Stoecker, ``{Canonical statistical model
  analysis of p-p , p -Pb, and Pb-Pb collisions at energies available at the
  CERN Large Hadron Collider}'',
  \href{http://dx.doi.org/10.1103/PhysRevC.100.054906}{{\em Phys. Rev. C}
  {\bfseries 100} (2019) 054906},
  \href{http://arxiv.org/abs/1906.03145}{{\ttfamily arXiv:1906.03145
  [hep-ph]}}.

\bibitem{ALICE:2020jsh}
{\bfseries ALICE} Collaboration, S.~Acharya {\em et~al.}, ``{Production of
  light-flavor hadrons in pp collisions at $\sqrt{s}~=~7\text { and }\sqrt{s} =
  13 \, \text { TeV} $}'',
  \href{http://dx.doi.org/10.1140/epjc/s10052-020-08690-5}{{\em Eur. Phys. J.
  C} {\bfseries 81} (2021) 256},
  \href{http://arxiv.org/abs/2005.11120}{{\ttfamily arXiv:2005.11120
  [nucl-ex]}}.

\bibitem{Cronin:1974zm}
J.~W. Cronin, H.~J. Frisch, M.~J. Shochet, J.~P. Boymond, R.~Mermod, P.~A.
  Piroue, and R.~L. Sumner, ``{Production of hadrons with large transverse
  momentum at 200, 300, and 400 GeV}'',
  \href{http://dx.doi.org/10.1103/PhysRevD.11.3105}{{\em Phys. Rev. D}
  {\bfseries 11} (1975) 3105--3123}.

\bibitem{ALICE:2021rpa}
{\bfseries ALICE} Collaboration, S.~Acharya {\em et~al.},
  ``{$\mathrm{K}^{*}$(892)$^{0}$ and $\phi$(1020) production in p--Pb
  collisions at $\sqrt{s_{\mathrm{NN}}} =$~8.16~TeV}'',
  \href{http://dx.doi.org/10.1103/PhysRevC.107.055201}{{\em Phys. Rev. C}
  {\bfseries 107} (2023) 055201},
  \href{http://arxiv.org/abs/2110.10042}{{\ttfamily arXiv:2110.10042
  [nucl-ex]}}.

\end{thebibliography}\endgroup

\newpage
\appendix

\section{The ALICE Collaboration}
\label{app:collab}
\begin{flushleft} 
\small

S.~Acharya\,\orcidlink{0000-0002-9213-5329}\,$^{\rm 128}$, 
D.~Adamov\'{a}\,\orcidlink{0000-0002-0504-7428}\,$^{\rm 87}$, 
G.~Aglieri Rinella\,\orcidlink{0000-0002-9611-3696}\,$^{\rm 33}$, 
M.~Agnello\,\orcidlink{0000-0002-0760-5075}\,$^{\rm 30}$, 
N.~Agrawal\,\orcidlink{0000-0003-0348-9836}\,$^{\rm 52}$, 
Z.~Ahammed\,\orcidlink{0000-0001-5241-7412}\,$^{\rm 136}$, 
S.~Ahmad\,\orcidlink{0000-0003-0497-5705}\,$^{\rm 16}$, 
S.U.~Ahn\,\orcidlink{0000-0001-8847-489X}\,$^{\rm 72}$, 
I.~Ahuja\,\orcidlink{0000-0002-4417-1392}\,$^{\rm 38}$, 
A.~Akindinov\,\orcidlink{0000-0002-7388-3022}\,$^{\rm 142}$
% ...

\section*{Affiliation Notes}

$^{\rm I}$ Deceased\\
$^{\rm II}$ Also at: Max-Planck-Institut fur Physik, Munich, Germany\\
$^{\rm III}$ Also at: Italian National Agency for New Technologies, Energy and Sustainable Economic Development (ENEA), Bologna, Italy\\
$^{\rm IV}$ Also at: Dipartimento DET del Politecnico di Torino, Turin, Italy\\
$^{\rm V}$ Also at: Yildiz Technical University, Istanbul, T\"{u}rkiye\\
$^{\rm VI}$ Also at: Department of Applied Physics, Aligarh Muslim University, Aligarh, India\\
$^{\rm VII}$ Also at: Institute of Theoretical Physics, University of Wroclaw, Poland\\
$^{\rm VIII}$ Also at: An institution covered by a cooperation agreement with CERN\\

\section*{Collaboration Institutes}

$^{1}$ A.I. Alikhanyan National Science Laboratory (Yerevan Physics Institute) Foundation, Yerevan, Armenia\\
$^{2}$ AGH University of Krakow, Cracow, Poland\\
$^{3}$ Bogolyubov Institute for Theoretical Physics, National Academy of Sciences of Ukraine, Kiev, Ukraine\\
$^{4}$ Bose Institute, Department of Physics  and Centre for Astroparticle Physics and Space Science (CAPSS), Kolkata, India\\
$^{5}$ California Polytechnic State University, San Luis Obispo, California, United States\\
$^{6}$ Central China Normal University, Wuhan, China\\
$^{7}$ Centro de Aplicaciones Tecnol\'{o}gicas y Desarrollo Nuclear (CEADEN), Havana, Cuba\\
$^{8}$ Centro de Investigaci\'{o}n y de Estudios Avanzados (CINVESTAV), Mexico City and M\'{e}rida, Mexico\\
$^{9}$ Chicago State University, Chicago, Illinois, United States\\
$^{10}$ China Institute of Atomic Energy, Beijing, China\\
$^{11}$ China University of Geosciences, Wuhan, China\\
$^{12}$ Chungbuk National University, Cheongju, Republic of Korea\\
$^{13}$ Comenius University Bratislava, Faculty of Mathematics, Physics and Informatics, Bratislava, Slovak Republic\\
$^{14}$ COMSATS University Islamabad, Islamabad, Pakistan\\
$^{15}$ Creighton University, Omaha, Nebraska, United States\\
$^{16}$ Department of Physics, Aligarh Muslim University, Aligarh, India\\
$^{17}$ Department of Physics, Pusan National University, Pusan, Republic of Korea\\
$^{18}$ Department of Physics, Sejong University, Seoul, Republic of Korea\\
$^{19}$ Department of Physics, University of California, Berkeley, California, United States\\
$^{20}$ Department of Physics, University of Oslo, Oslo, Norway\\
$^{21}$ Department of Physics and Technology, University of Bergen, Bergen, Norway\\
$^{22}$ Dipartimento di Fisica, Universit\`{a} di Pavia, Pavia, Italy\\
$^{23}$ Dipartimento di Fisica dell'Universit\`{a} and Sezione INFN, Cagliari, Italy\\
$^{24}$ Dipartimento di Fisica dell'Universit\`{a} and Sezione INFN, Trieste, Italy\\
$^{25}$ Dipartimento di Fisica dell'Universit\`{a} and Sezione INFN, Turin, Italy\\
$^{26}$ Dipartimento di Fisica e Astronomia dell'Universit\`{a} and Sezione INFN, Bologna, Italy\\
$^{27}$ Dipartimento di Fisica e Astronomia dell'Universit\`{a} and Sezione INFN, Catania, Italy\\
$^{28}$ Dipartimento di Fisica e Astronomia dell'Universit\`{a} and Sezione INFN, Padova, Italy\\
$^{29}$ Dipartimento di Fisica `E.R.~Caianiello' dell'Universit\`{a} and Gruppo Collegato INFN, Salerno, Italy\\
$^{30}$ Dipartimento DISAT del Politecnico and Sezione INFN, Turin, Italy\\
$^{31}$ Dipartimento di Scienze MIFT, Universit\`{a} di Messina, Messina, Italy\\
$^{32}$ Dipartimento Interateneo di Fisica `M.~Merlin' and Sezione INFN, Bari, Italy\\
$^{33}$ European Organization for Nuclear Research (CERN), Geneva, Switzerland\\
$^{34}$ Faculty of Electrical Engineering, Mechanical Engineering and Naval Architecture, University of Split, Split, Croatia\\
$^{35}$ Faculty of Engineering and Science, Western Norway University of Applied Sciences, Bergen, Norway\\
$^{36}$ Faculty of Nuclear Sciences and Physical Engineering, Czech Technical University in Prague, Prague, Czech Republic\\
$^{37}$ Faculty of Physics, Sofia University, Sofia, Bulgaria\\
$^{38}$ Faculty of Science, P.J.~\v{S}af\'{a}rik University, Ko\v{s}ice, Slovak Republic\\
$^{39}$ Frankfurt Institute for Advanced Studies, Johann Wolfgang Goethe-Universit\"{a}t Frankfurt, Frankfurt, Germany\\
$^{40}$ Fudan University, Shanghai, China\\
$^{41}$ Gangneung-Wonju National University, Gangneung, Republic of Korea\\
$^{42}$ Gauhati University, Department of Physics, Guwahati, India\\
$^{43}$ Helmholtz-Institut f\"{u}r Strahlen- und Kernphysik, Rheinische Friedrich-Wilhelms-Universit\"{a}t Bonn, Bonn, Germany\\
$^{44}$ Helsinki Institute of Physics (HIP), Helsinki, Finland\\
$^{45}$ High Energy Physics Group,  Universidad Aut\'{o}noma de Puebla, Puebla, Mexico\\
$^{46}$ Horia Hulubei National Institute of Physics and Nuclear Engineering, Bucharest, Romania\\
$^{47}$ HUN-REN Wigner Research Centre for Physics, Budapest, Hungary\\
$^{48}$ Indian Institute of Technology Bombay (IIT), Mumbai, India\\
$^{49}$ Indian Institute of Technology Indore, Indore, India\\
$^{50}$ INFN, Laboratori Nazionali di Frascati, Frascati, Italy\\
$^{51}$ INFN, Sezione di Bari, Bari, Italy\\
$^{52}$ INFN, Sezione di Bologna, Bologna, Italy\\
$^{53}$ INFN, Sezione di Cagliari, Cagliari, Italy\\
$^{54}$ INFN, Sezione di Catania, Catania, Italy\\
$^{55}$ INFN, Sezione di Padova, Padova, Italy\\
$^{56}$ INFN, Sezione di Pavia, Pavia, Italy\\
$^{57}$ INFN, Sezione di Torino, Turin, Italy\\
$^{58}$ INFN, Sezione di Trieste, Trieste, Italy\\
$^{59}$ Inha University, Incheon, Republic of Korea\\
$^{60}$ Institute for Gravitational and Subatomic Physics (GRASP), Utrecht University/Nikhef, Utrecht, Netherlands\\
$^{61}$ Institute of Experimental Physics, Slovak Academy of Sciences, Ko\v{s}ice, Slovak Republic\\
$^{62}$ Institute of Physics, Homi Bhabha National Institute, Bhubaneswar, India\\
$^{63}$ Institute of Physics of the Czech Academy of Sciences, Prague, Czech Republic\\
$^{64}$ Institute of Space Science (ISS), Bucharest, Romania\\
$^{65}$ Institut f\"{u}r Kernphysik, Johann Wolfgang Goethe-Universit\"{a}t Frankfurt, Frankfurt, Germany\\
$^{66}$ Instituto de Ciencias Nucleares, Universidad Nacional Aut\'{o}noma de M\'{e}xico, Mexico City, Mexico\\
$^{67}$ Instituto de F\'{i}sica, Universidade Federal do Rio Grande do Sul (UFRGS), Porto Alegre, Brazil\\
$^{68}$ Instituto de F\'{\i}sica, Universidad Nacional Aut\'{o}noma de M\'{e}xico, Mexico City, Mexico\\
$^{69}$ iThemba LABS, National Research Foundation, Somerset West, South Africa\\
$^{70}$ Jeonbuk National University, Jeonju, Republic of Korea\\
$^{71}$ Johann-Wolfgang-Goethe Universit\"{a}t Frankfurt Institut f\"{u}r Informatik, Fachbereich Informatik und Mathematik, Frankfurt, Germany\\
$^{72}$ Korea Institute of Science and Technology Information, Daejeon, Republic of Korea\\
$^{73}$ KTO Karatay University, Konya, Turkey\\
$^{74}$ Laboratoire de Physique Subatomique et de Cosmologie, Universit\'{e} Grenoble-Alpes, CNRS-IN2P3, Grenoble, France\\
$^{75}$ Lawrence Berkeley National Laboratory, Berkeley, California, United States\\
$^{76}$ Lund University Department of Physics, Division of Particle Physics, Lund, Sweden\\
$^{77}$ Nagasaki Institute of Applied Science, Nagasaki, Japan\\
$^{78}$ Nara Women{'}s University (NWU), Nara, Japan\\
$^{79}$ National and Kapodistrian University of Athens, School of Science, Department of Physics , Athens, Greece\\
$^{80}$ National Centre for Nuclear Research, Warsaw, Poland\\
$^{81}$ National Institute of Science Education and Research, Homi Bhabha National Institute, Jatni, India\\
$^{82}$ National Nuclear Research Center, Baku, Azerbaijan\\
$^{83}$ National Research and Innovation Agency - BRIN, Jakarta, Indonesia\\
$^{84}$ Niels Bohr Institute, University of Copenhagen, Copenhagen, Denmark\\
$^{85}$ Nikhef, National institute for subatomic physics, Amsterdam, Netherlands\\
$^{86}$ Nuclear Physics Group, STFC Daresbury Laboratory, Daresbury, United Kingdom\\
$^{87}$ Nuclear Physics Institute of the Czech Academy of Sciences, Husinec-\v{R}e\v{z}, Czech Republic\\
$^{88}$ Oak Ridge National Laboratory, Oak Ridge, Tennessee, United States\\
$^{89}$ Ohio State University, Columbus, Ohio, United States\\
$^{90}$ Physics department, Faculty of science, University of Zagreb, Zagreb, Croatia\\
$^{91}$ Physics Department, Panjab University, Chandigarh, India\\
$^{92}$ Physics Department, University of Jammu, Jammu, India\\
$^{93}$ Physics Program and International Institute for Sustainability with Knotted Chiral Meta Matter (SKCM2), Hiroshima University, Hiroshima, Japan\\
$^{94}$ Physikalisches Institut, Eberhard-Karls-Universit\"{a}t T\"{u}bingen, T\"{u}bingen, Germany\\
$^{95}$ Physikalisches Institut, Ruprecht-Karls-Universit\"{a}t Heidelberg, Heidelberg, Germany\\
$^{96}$ Physik Department, Technische Universit\"{a}t M\"{u}nchen, Munich, Germany\\
$^{97}$ Politecnico di Bari and Sezione INFN, Bari, Italy\\
$^{98}$ Research Division and ExtreMe Matter Institute EMMI, GSI Helmholtzzentrum f\"ur Schwerionenforschung GmbH, Darmstadt, Germany\\
$^{99}$ Saga University, Saga, Japan\\
$^{100}$ Saha Institute of Nuclear Physics, Homi Bhabha National Institute, Kolkata, India\\
$^{101}$ School of Physics and Astronomy, University of Birmingham, Birmingham, United Kingdom\\
$^{102}$ Secci\'{o}n F\'{\i}sica, Departamento de Ciencias, Pontificia Universidad Cat\'{o}lica del Per\'{u}, Lima, Peru\\
$^{103}$ Stefan Meyer Institut f\"{u}r Subatomare Physik (SMI), Vienna, Austria\\
$^{104}$ SUBATECH, IMT Atlantique, Nantes Universit\'{e}, CNRS-IN2P3, Nantes, France\\
$^{105}$ Sungkyunkwan University, Suwon City, Republic of Korea\\
$^{106}$ Suranaree University of Technology, Nakhon Ratchasima, Thailand\\
$^{107}$ Technical University of Ko\v{s}ice, Ko\v{s}ice, Slovak Republic\\
$^{108}$ The Henryk Niewodniczanski Institute of Nuclear Physics, Polish Academy of Sciences, Cracow, Poland\\
$^{109}$ The University of Texas at Austin, Austin, Texas, United States\\
$^{110}$ Universidad Aut\'{o}noma de Sinaloa, Culiac\'{a}n, Mexico\\
$^{111}$ Universidade de S\~{a}o Paulo (USP), S\~{a}o Paulo, Brazil\\
$^{112}$ Universidade Estadual de Campinas (UNICAMP), Campinas, Brazil\\
$^{113}$ Universidade Federal do ABC, Santo Andre, Brazil\\
$^{114}$ Universitatea Nationala de Stiinta si Tehnologie Politehnica Bucuresti, Bucharest, Romania\\
$^{115}$ University of Cape Town, Cape Town, South Africa\\
$^{116}$ University of Derby, Derby, United Kingdom\\
$^{117}$ University of Houston, Houston, Texas, United States\\
$^{118}$ University of Jyv\"{a}skyl\"{a}, Jyv\"{a}skyl\"{a}, Finland\\
$^{119}$ University of Kansas, Lawrence, Kansas, United States\\
$^{120}$ University of Liverpool, Liverpool, United Kingdom\\
$^{121}$ University of Science and Technology of China, Hefei, China\\
$^{122}$ University of South-Eastern Norway, Kongsberg, Norway\\
$^{123}$ University of Tennessee, Knoxville, Tennessee, United States\\
$^{124}$ University of the Witwatersrand, Johannesburg, South Africa\\
$^{125}$ University of Tokyo, Tokyo, Japan\\
$^{126}$ University of Tsukuba, Tsukuba, Japan\\
$^{127}$ Universit\"{a}t M\"{u}nster, Institut f\"{u}r Kernphysik, M\"{u}nster, Germany\\
$^{128}$ Universit\'{e} Clermont Auvergne, CNRS/IN2P3, LPC, Clermont-Ferrand, France\\
$^{129}$ Universit\'{e} de Lyon, CNRS/IN2P3, Institut de Physique des 2 Infinis de Lyon, Lyon, France\\
$^{130}$ Universit\'{e} de Strasbourg, CNRS, IPHC UMR 7178, F-67000 Strasbourg, France, Strasbourg, France\\
$^{131}$ Universit\'{e} Paris-Saclay, Centre d'Etudes de Saclay (CEA), IRFU, D\'{e}partment de Physique Nucl\'{e}aire (DPhN), Saclay, France\\
$^{132}$ Universit\'{e}  Paris-Saclay, CNRS/IN2P3, IJCLab, Orsay, France\\
$^{133}$ Universit\`{a} degli Studi di Foggia, Foggia, Italy\\
$^{134}$ Universit\`{a} del Piemonte Orientale, Vercelli, Italy\\
$^{135}$ Universit\`{a} di Brescia, Brescia, Italy\\
$^{136}$ Variable Energy Cyclotron Centre, Homi Bhabha National Institute, Kolkata, India\\
$^{137}$ Warsaw University of Technology, Warsaw, Poland\\
$^{138}$ Wayne State University, Detroit, Michigan, United States\\
$^{139}$ Yale University, New Haven, Connecticut, United States\\
$^{140}$ Yonsei University, Seoul, Republic of Korea\\
$^{141}$  Zentrum  f\"{u}r Technologie und Transfer (ZTT), Worms, Germany\\
$^{142}$ Affiliated with an institute covered by a cooperation agreement with CERN\\
$^{143}$ Affiliated with an international laboratory covered by a cooperation agreement with CERN.\\

\end{flushleft} 

\end{document}